\documentclass[12pt]{article}

\usepackage{sbc-template}

\usepackage{graphicx}
\usepackage{listings}
\usepackage[T1]{fontenc}
\usepackage{alltt, moreverb}	
\usepackage{lipsum}		
\usepackage{currfile}		
\usepackage[final]{pdfpages}	
\usepackage{longtable}		
\usepackage{tabularx}      
\usepackage{lscape}
\usepackage[none]{hyphenat}
\usepackage{makecell}
\usepackage{enumitem} 
\usepackage{array}
\usepackage{multirow}
\usepackage{xurl}
\usepackage{tablefootnote}
\usepackage{xltabular}
\usepackage{longtable}
\usepackage{booktabs}
\usepackage{pdflscape}
\usepackage{pdfpages}
\usepackage{afterpage}
\usepackage{ragged2e}
\usepackage{fancyhdr}
\usepackage{quoting}
\usepackage{mdframed}
\usepackage{float}
\usepackage{ulem}
\usepackage{lscape}
\usepackage{diagbox}
\usepackage{threeparttable}
\usepackage{accessibility}
\usepackage{ragged2e}
\usepackage{colortbl}
\usepackage{natbib}
\renewcommand{\cite}{\citep}
\PassOptionsToPackage{hyphens}{url}\usepackage{hyperref}
     
\title{Decoding the Sociotechnical Dimensions of Digital Misinformation: A Comprehensive Literature Review}

\author{Alisson Andrey Puska, Luiz Adolpho Baroni, Roberto Pereira}
\address{Human-Computer Interaction Laboratory, \\ Federal University of Paraná (UFPR) \\ Curitiba, Paraná, Brazil 
\email{alisson.puska, sirLouiz@gmail.com; rpereira@inf.ufpr.br}}

\begin{document}

\maketitle

\begin{abstract}
This paper presents a systematic literature review in Computer Science that provide an overview of the initiatives related to digital misinformation. This is an exploratory study that covers research from 1993 to 2020, focusing on the investigation of the phenomenon of misinformation. The review consists of 788 studies from SCOPUS, IEEE, and ACM digital libraries, synthesizing the primary research directions and sociotechnical challenges. These challenges are classified into Physical, Empirical, Syntactic, Semantic, Pragmatic, and Social dimensions, drawing from Organizational Semiotics. The mapping identifies issues related to the concept of misinformation, highlights deficiencies in mitigation strategies, discusses challenges in approaching stakeholders, and unveils various sociotechnical aspects relevant to understanding and mitigating the harmful effects of digital misinformation. As contributions, this study present a novel categorization of mitigation strategies, a sociotechnical taxonomy for classifying types of false information and elaborate on the inter-relation of sociotechnical aspects and their impacts.
\end{abstract}

\providecommand{\keywords}[1]
{
  \small	
  \textbf{\textit{Keywords---}} #1
}


\section{Introduction}

In today's digital landscape, where personal beliefs and emotional resonances often eclipse empirical facts in shaping perceptions~\citep{flood2016post}, the proliferation of misleading information has become a critical societal challenge. Particularly within online environments, such misinformation can amplify misconceptions about real-world events, influencing individual behaviors and decisions with far-reaching consequences. From economic repercussions, exemplified by stock market instabilities~\citep{hwang2012socialbots}, to public health crises, underscored by vaccine skepticism~\citep{vraga2018not, lenzer2011fake}, the dissemination of spurious data poses tangible risks across various societal domains~\citep{rowe2004two, rowe2006taxonomy}.

Distinguishing between disinformation and misinformation is crucial for understanding and addressing the spread of digital falsehoods. Disinformation, defined as deliberately incorrect information designed to deceive, manipulate beliefs, or induce decision errors, contrasts with misinformation, which misleads without intentional deceit \citep{tudjman2003information}. This differentiation informs the development of targeted mitigation strategies and underscores the complexity of countering digital false information, as it highlights the varying motivations behind the spread of digital falsehoods and the complexities involved in combating them\footnote{
In the remainder of the article, the term "digital false information" will be used to refer to both types.}.

The scientific community is deeply involved in exploring and counteracting the challenges posed by the spread of digital false information, with efforts covering both technical and social aspects. This includes studying how false information spreads across networks~\cite{shao2017spread}, the human behaviors that influence its consumption~\cite{shrestha2019online, raman2019manipulating}, and the linguistic characteristics of deceptive messages to create automated countermeasures~\cite{perez2017automatic}. These efforts underscore the importance of recognizing and examining the sociotechnical elements at various levels to fully understand and address the impact of false information.  

Recognizing the need for a holistic grasp of the research within the Computer Science domain, we advocate for a thorough and cross-cutting exploration of the literature through a systematic mapping study. Such an endeavor is essential to attain a comprehensive understanding of the ongoing efforts in tackling the challenges posed by digital false information, thereby shedding light on how the intricate sociotechnical aspects of this phenomenon are being effectively addressed.

We present a systematic literature review analyzing 788 studies from SCOPUS, IEEE, and ACM digital libraries, spanning from 1993 to 2020. Our review not only synthesizes the primary research directions but also introduces a novel framework for understanding the sociotechnical challenges posed by digital false information. These challenges are categorized into six dimensions—Physical, Empirical, Syntactic, Semantic, Pragmatic, and Social—drawing from Organizational Semiotics\cite{stamper1991semiotic}, and underscore the need for an integrated approach that marries technical innovation with an understanding of social dynamics.

The findings underscore several key areas for future research and development: 
\vspace{-0.2in}
\begin{itemize}[noitemsep]
    \item Expanding Methodological Approaches: The review points out the need for more comprehensive and multidisciplinary approaches beyond linguistic analysis for tackling digital false information. Expanding methodologies to include more diverse datasets and considering factors beyond the text, such as multimedia content and user behavior, could provide a more holistic understanding of digital false information phenomena.

    \item Improving Dataset Diversity:  There is a critique of the limited availability and representation within datasets, particularly regarding the scarcity of diverse datasets covering various topics and languages. Efforts to include or create more varied datasets could significantly improve the detection and analysis of digital false information across different contexts and cultures.

    \item Enhancing Stakeholder Engagement: The findings show that the involvement of companies, governments, and users is limited in current mitigation strategies. Developing frameworks for greater collaboration and engagement with these stakeholders could lead to more effective and implementable solutions.

    \item Prioritizing Ethical and Responsible Solutions: The results highlight concerns about potentially invasive and harmful persuasive technologies used to combat digital false information. There is a call for more discussion on the ethical design and implementation of solutions, ensuring they do not inadvertently harm users or infringe on their rights.

    \item Deepening Sociotechnical Insights: we elaborate on the complex interrelations between the sociotechnical aspects of digital false information. A more thorough exploration of the interconnections between social and technical factors influencing digital false information spread and reception is encouraged, to identify effective mitigation strategies.

    \item Addressing Research Fragmentation: The review identifies a tendency to segment research into specific stages of digital false information without considering the lifecycle as a whole. Future research could strive for a more integrated approach that addresses the continuum of digital false information from creation to consumption and its impact.
    
    \item Broadening Cultural and Linguistic Perspectives: The dominance of English-language datasets and research is noted as a limitation. Expanding research to include more languages and cultural contexts would not only enhance the understanding of digital false information globally but also improve the development of situated mitigation strategies.

\end{itemize}

This paper's contributions are manifold, offering a systematic review of the digital false information landscape, a novel categorization of sociotechnical challenges, and an in-depth analysis of mitigation strategies and digital false information types. By elucidating the intricate sociotechnical interrelations within digital falsehoods, this research provides a foundational framework for future investigations and the development of more nuanced, effective countermeasures.

\section{Motivation to conduct a systematic review}
\label{proto}
\begin{figure}[]
\centering
\includegraphics[scale=0.39]{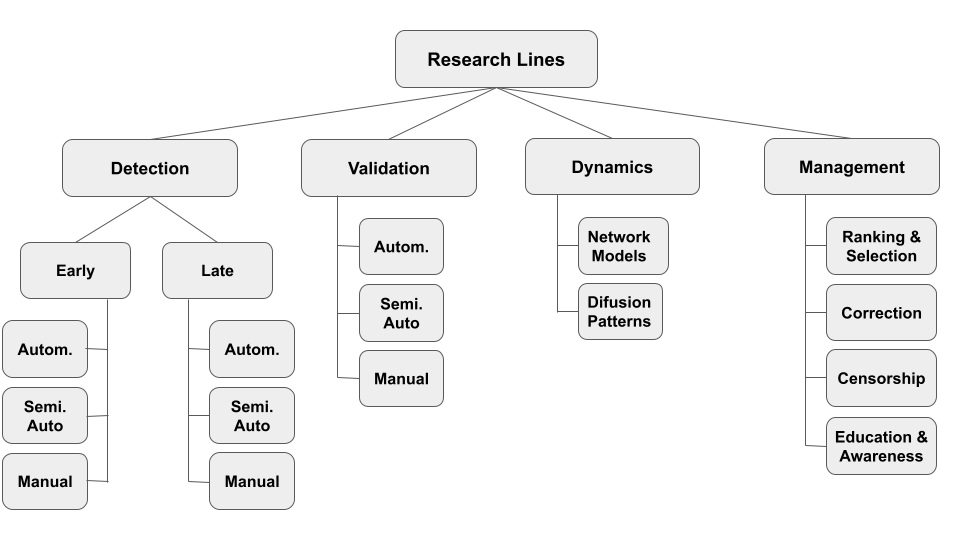}
\caption{Research lines on digital false information.}
\label{Fig:art}
\end{figure}

The inception of our research was driven by an exploratory study aimed at scrutinizing literature reviews and systematic mappings focused on digital false information from 2010 to 2020. Our goal was to aggregate a holistic view of the challenges and methodologies applied in the study and mitigation of digital false information. We identified 39 literature reviews that span a variety of digital false information forms. These served as foundational references for coding, structuring, and dissecting the identified challenges and proposed solutions within the field. This preliminary analysis delineated four primary research streams—Detection, Validation, Dynamics, and Management—each thoroughly documented in existing scholarship \cite{almaliki2019online, fernandez2018online}. Figure~\ref{Fig:art} visually synthesizes how these areas have been represented in prior works \cite{almaliki2019misinformation, fernandez2018online}.

Detection efforts are predicated on identifying digital false information through content examination, user profiling, and social network analysis. Validation focuses on verifying content accuracy, while Dynamics explores the patterns of digital false information spread across social networks. Management strategies are aimed at developing countermeasures, such as optimizing how corrections are presented or educating audiences on identifying false information.

Our review reveals that detection can occur either "early" before digital false information widely spreads, employing techniques like machine-mediated recognition of malicious accounts \cite{pal2016reviewing}, or "late", identifying widespread digital false information through methods like manual fact-checking \cite{saquete2020fighting}. Validation employs diverse techniques to ascertain the veracity of information, including machine learning and natural language processing \cite{bondielli2019survey}.

The study of Dynamics has gained attention for its focus on social network structures and digital false information spread patterns, aiming to characterize and mitigate the dissemination of false information \cite{rana2019rumor, almaliki2019online}. Management strategies encompass a broad range of approaches, from classifying digital false information to educating the public on discerning truth from falsehood \cite{manzoor2019fake, pal2019communicating}.

\subsection{Problems and Challenges}
Our examination highlights several deficiencies within these research streams. The identified challenges in the reviews, detailed in Table~\ref{tab:prblms}, point to a need for more comprehensive and integrated approaches that consider the multifaceted nature of digital false information. These include the necessity for diverse datasets, methodological innovation, and a greater focus on the sociotechnical dimensions of digital false information.

\begin{table}[]
\centering
\footnotesize
\caption{Challenges found in the literature}
\label{tab:prblms}
\renewcommand{\tabularxcolumn}[1]{m{#1}}
\begin{tabularx}{\textwidth}{>{\hsize=0.35\hsize\linewidth=\hsize\raggedright\arraybackslash}X>{\hsize=1.35\hsize\linewidth=\hsize\raggedright\arraybackslash}X>{\hsize=1.3\hsize\linewidth=\hsize\raggedright\arraybackslash}X}
\hline
Research Line & Complexity & Multidisciplinarity \\
\hline
Detection &
Limited understanding of the phenomenon~\cite{fernandez2018online};
Data and methodology limitations\cite{zhang2020overview, lozano2020veracity, al2019deep};
Emphasis on technical solutions\cite{fernandez2018online};
limited engagement with diverse content themes~\cite{habib2019false};
&
Focus on technical aspects\cite{fernandez2018online, saquete2020fighting};
Addressing pre-determined sets of sociotechnical aspects\cite{jindal2020newsbag};
Credibility attributes based on generic models\cite{saquete2020fighting};
\\ \hline
Validation &
Dependent on verified data and reliable sources\cite{al2019deep};
Corrections distant from the user\cite{fernandez2018online};
Limited access to labeled databases\cite{zhang2020overview};
Tendency to focus on detection rather than validation \textit{per se}\cite{al2019deep};
Limited scope regarding content themes\cite{al2019deep};
&
Technical aspects emphasized\cite{fernandez2018online};
Linguistic aspects emphasized\cite{saquete2020fighting};
Credibility attributes based on generic models from other media\cite{saquete2020fighting};
\\ \hline
Dynamics &
Prevalence of focus on topology\cite{rana2019rumor};
Biology-inspired models (viral propagation) less representative\cite{fernandez2018online};
&
Technical aspects emphasized\cite{fernandez2018online};
Social aspects focus on socio-demographic attributes\cite{rana2019rumor};
Emphasis on motivations and relationships of stakeholders\cite{fernandez2018online};
\\ \hline
Management &
Focus on generating and disseminating corrections\cite{fernandez2018online};
Fragmented strategies, with little integration of solutions\cite{caulfield2019jenny};
Predominance of control tasks (identification and censorship)\cite{shelke2019source};
&
Technical aspects emphasized\cite{fernandez2018online};
User-centered approaches lacking\cite{fernandez2018online};
Need for exploratory studies on human-factors, such as cognitive biases~\cite{al2019deep};
\\\hline
\end{tabularx}
\end{table}

The quest for effective solutions necessitates a shift towards more multidisciplinary (even transdisciplinary) methodologies that transcend mere content analysis, encouraging the integration of technical, human, and organizational perspectives. This approach not only addresses the detection and validation of digital false information but also considers its dynamic spread and the development of comprehensive management strategies.

\subsection{Discussion}

The exploration of digital false information through the lens of a systematic review has illuminated the necessity for a multifaceted approach to understanding and mitigating its impact. Our initial foray into the literature from 2010 to 2020 has delineated the primary research streams of Detection, Validation, Dynamics, and Management, each presenting unique challenges and necessitating diverse methodologies for comprehensive analysis. The identified research streams and their associated challenges underline the complexity of digital false information and the importance of adopting integrated sociotechnical methodologies for effective management and mitigation.

Our analysis reveals a critical gap in current research methodologies, which often prioritize technical solutions at the expense of a broader, more holistic understanding of digital false information. This oversight limits the potential for developing effective countermeasures that address not only the technological aspects of digital false information but also the human and organizational factors that play a pivotal role in its spread and impact. The challenges outlined in our review, particularly those related to the need for diverse datasets, methodological innovation, and a focus on sociotechnical dimensions, signal a call for a paradigm shift towards research methodologies that embrace the complexity of digital false information.

The discussion of these findings points towards the importance of adopting a sociotechnical perspective that integrates technical, human, social and organizational aspects in the study and management of digital false information. By emphasizing the need for multidisciplinary approaches, our analysis advocates for research that not only detects and validates digital false information but also understands its dynamics and develops comprehensive strategies for management. This approach recognizes the intertwined nature of technology and society, highlighting the significance of considering the human factors, network structures, and organizational contexts that influence the spread and reception of digital false information.

Furthermore, the identification of fragmented strategies and the predominance of technical approaches in existing literature underscore the necessity for more inclusive research that considers the diverse stakeholders involved in the digital false information ecosystem. By broadening the scope of investigation to include a wider range of sociotechnical aspects, research can better identify and address the underlying causes of digital false information, its propagation mechanisms, and the most effective strategies for combating its spread.

In conclusion, our discussion emphasizes the need for a systematic mapping study that extends beyond the current temporal and thematic limitations, aiming to capture a comprehensive view of how digital false information is addressed across different contexts. Through such an endeavor, we aspire to uncover novel insights and frameworks that will inform future research and practice, ultimately contributing to more effective solutions in the fight against digital falsehoods. This systematic approach will not only enrich our understanding of the phenomenon but also pave the way for the development of more nuanced and integrated strategies for managing the complexities of digital false information in our increasingly interconnected world.

\section{Methodological Framework for Systematic Literature Mapping}
\label{Histo}

\urldef{\footurl}\url{https://docs.google.com/spreadsheets/d/1lZORTRirkr-kKYzZXC9gV8lip3d31fsVG7X8nY7RLbg/}

The mapping was proposed to obtain an updated and comprehensive overview of research on the phenomenon of digital false information from the perspective of the Computer Science field, within the context of the digital libraries of ACM, Scopus, and IEEE. The search key was based on the keywords identified in the exploratory study (Table~\ref{string}). The mapping covered the period from 1993\footnote{The year of the NCSA Mosaic's inception, which significantly contributed to the popularization of the Web~\cite{schatz1994ncsa}. The main author considers this as a milestone for the social aspects of the digital false information phenomenon.} until 2020.

\begin{table}[h]
\centering
\caption{Search Key Base}
\label{string}
\def\arraystretch{1.2} 
\begin{tabular}{c@{} p{4in} c@{}}
\toprule
& (\textquotedbl disinformation\textquotedbl{} OR \textquotedbl misinformation\textquotedbl{} OR \textquotedbl fake news\textquotedbl{} OR \textquotedbl false news\textquotedbl{} OR \textquotedbl fabricated news\textquotedbl{} OR \textquotedbl hoax\textquotedbl{} OR \textquotedbl rumor\textquotedbl{} OR \textquotedbl false information\textquotedbl{} OR \textquotedbl fake information\textquotedbl{} OR \textquotedbl fabricated information\textquotedbl{}) \\
\bottomrule
\end{tabular}
\end{table}


Given the limitations and gaps identified in the exploratory phase, we formulated four critical Research Questions (RQs) to guide our mapping:

\begin{itemize}
\item QP01 - What are the types of digital false information studied?
\begin{itemize}
\item The exploratory study revealed divergences in the concepts of "disinformation" and "misinformation" and in the characterization of types of false information. This research question aims to understand the issues with the concepts of misinformation and disinformation, identify the types of digital false information (satire, fake news, etc.), and the sociotechnical aspects that characterize them.
\end{itemize}
\item QP02 - Which and how stakeholders are addressed?
\begin{itemize}
\item The exploratory study showed a tendency to address obvious stakeholders (e.g., speaker and audience) while leaving other relevant actors out (e.g., digital platform controllers, service providers, governments). This research question aims to identify the addressed stakeholders and the methodology used to identify and characterize them.
\end{itemize}
\item QP03 - How do researchers address sociotechnical aspects?
\begin{itemize}
\item In the exploratory study, we observed a tendency to address technical aspects of digital false information and difficulties in considering human and social aspects in an integrated way. This research question aims to identify the scope of models and characterizations of digital false information cases considered by the solutions and how sociotechnical aspects are addressed.
\end{itemize}
\item QP04 - How do research approaches address the organization of digital false information case?
\begin{itemize}
\item The exploratory study revealed the tendency to address the phenomenon of digital false information with a focus on the consumption and dissemination of digital false information. This research question aims to identify the elements of a digital false information case and how studies address its organizational aspects.
\end{itemize}
\end{itemize}

\begin{figure}[]
\centering
\includegraphics[scale=0.22]{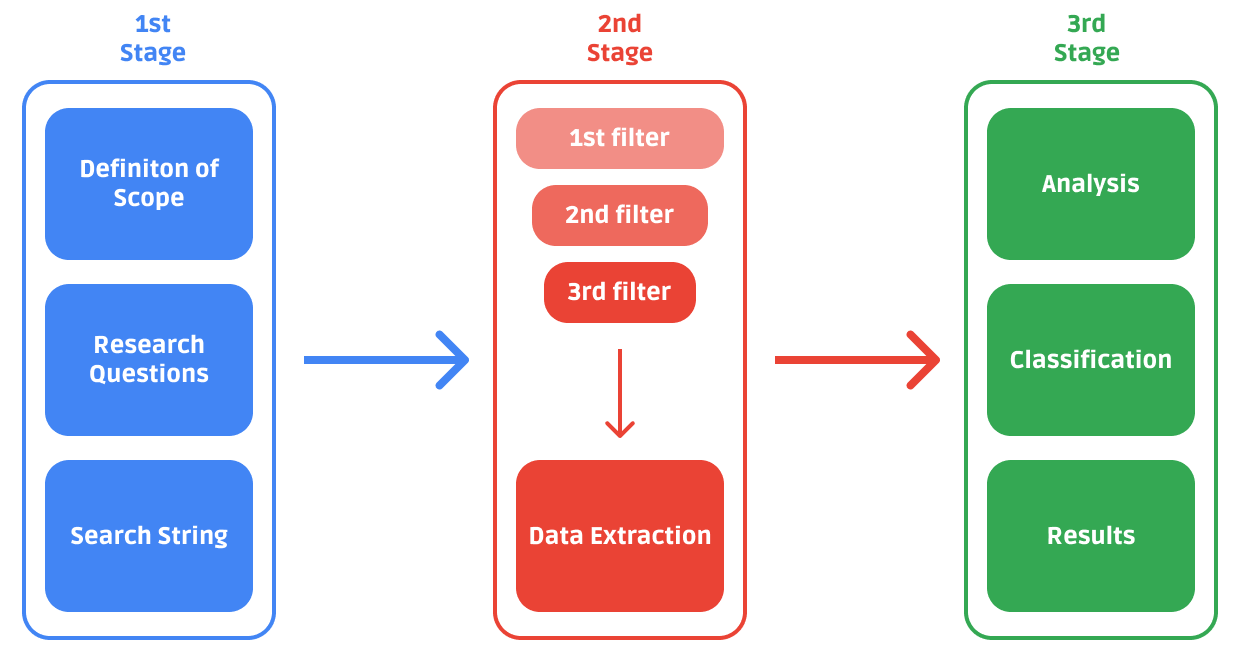}
\caption{Systematic Mapping Protocol}
\label{Fig:systematic}
\end{figure}

A team of three researchers from the Human-Computer Interaction laboratory at the Federal University of Paraná State in Brazil applied a meticulously designed three-stage mapping protocol to conduct this study, as illustrated in Figure~\ref{Fig:systematic}. The initial phase involved defining the scope, research questions, and the search strategy. The subsequent phase focused on filtering the gathered data and extracting relevant information. In the final phase, the team analyzed the data and engaged in discussions to derive the study's findings. The entire process spanned approximately one and a half years (January 2019 - June 2020).

\urldef{\footurlq}\url{https://docs.google.com/spreadsheets/d/1lZORTRirkr-kKYzZXC9gV8lip3d31fsVG7X8nY7RLbg/edit?usp=sharing}

\urldef{\footurlh}\url{https://docs.google.com/spreadsheets/d/1QqHWVMC_CMMwLePvYGf8miM7NlM-CxYz6Z1ttDi494Q/edit?usp=sharing/}

The data gathered were separated equally into three parts, each designated to one researcher. We used the Mendeley~\cite{BibEntry2020Apr} tool and Google Sheets for organizational reasons. The documents of the filtration phases\footnote{\footurlq} and the data extraction\footnote{\footurlh} were organized and synthesized in Google Sheets spreadsheet.

The research string returned 4946 titles. Mendeley automatically detected 355 duplicates. On the manual revision of each researcher, other 805 duplicates were found and removed during the filtering phases, resulting in a total of 1160 duplicates removed. The first filter revision process encompassed the reading of the abstract of each paper. It resulted in 1779 removed papers. After the first filter, the research team reviewed the results and discussed their differences, removing other 1766 papers. The total amount of removed titles from the first filter were 3545, including duplicates. The string of research included the term 'rumor', for having a large scope of definitions on the literature, which demanded attention on the revision, filtering, and discussion processes. 

The second filter encompassed the reading of the introduction of each paper. It resulted in 239 removed papers. After the second filter, the team reviewed the results, and divergences were discussed, resulting in a total of 258 papers removed (third filter). The consolidated list of papers selected for data extraction had 788 papers. Table~\ref{map:stat} summarizes the results of the filtration process.

\begin{table}[H]
\centering
\small
\caption{Statistics of the systematic mapping}
\label{map:stat}
\begin{tabular}{llll}
\hline
Statistics & Processed & Review & Total \\ \hline
All articles & & & 4591 \\
1st filter (excluded) & -1776 & -1733 & -3059 \\
2nd filter (excluded) & -275 & -19 & -294 \\
Data extracted & & & \textbf{788} \\ \hline
\end{tabular}
\end{table}

Each filtering procedure used a registration form for documentation purposes. Table~\ref{form_filters} shows the form structure. The team used a content analysis method~\cite{lazar2017research} to analyze the papers. At the end of each phase, the researchers discussed and reviewed the classification results. In the last phase, the main researcher analyzed a subset of the selected papers to extract information to answer the research questions. 

\begin{table}[h]
\centering
\caption{Filtering form}
\label{form_filters}
\begin{tabular}{llllll}
\textbf{ID}            & \textbf{Title}        & \textbf{Year}         & \textbf{Abstract}     & \textbf{Exclusion/Inclusion Criteria} & \textbf{Justification}     \\ \hline
\multicolumn{1}{|l|}{} & \multicolumn{1}{l|}{} & \multicolumn{1}{l|}{} & \multicolumn{1}{l|}{} & \multicolumn{1}{l|}{}                 & \multicolumn{1}{l|}{} \\ \hline
\end{tabular}
\end{table}

\section{Findings}

\subsection{Overview of Results: 1994 - 2020}

The popularization of social media and the growing consumption of information online in the 21st century raises concerns about the possible impacts of false information on a global scale~\cite{Cadwalladr2020, UN}. Recent events in health~\cite{G1, PolitiFact}, politics~\cite{Biden, Fatos}, economy~\cite{Martnnez2018Nov}, security, and civil rights~\cite{Martnnez2018Nov, BBCNews2019Apr} demonstrate the sinister potential of false information and its dissemination on social networks. 

In its first steps, the discussion still speculated that the Internet would find the tools to unravel any lie, especially the anonymity of the liar~\cite{neumann1996disinformation}. On the other hand, almost simultaneously, there were concerns about the potential for the Internet to become a vehicle for disinformation~\cite{Luciano1996, resnik1998medical}. In the current global scenario of online false information, the discrepancy between the tunes of each discussion indicates the complex challenge and diversity of ways to approach it.

\begin{figure}[h]
\centering
\includegraphics[scale=0.5]{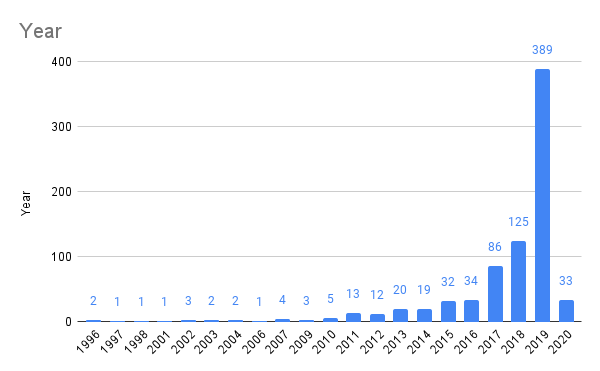}
\caption{Amount of publication by year}
\label{Fig:Pub_per_Year}
\end{figure}

Social media has clearly transformed the way people see and interact with each other. False information in the forms of lies, mistakes, or satires is as ubiquitous as digital communication itself\cite{lemieux2018leveraging}. Given the complexity of the false information phenomena online, we found in the literature a diversity of perspectives to approach the subject. Figure~\ref{Fig:Pub_per_Year} present an overview of the number of publications by year, indicating the growing research effort on the phenomena.

In the realm of computer science research on "digital false information," one of our initial challenges was the absence of a consensus in taxonomy~\cite{fard2019assessing, jiang2018linguistic}. This lack of agreement led to the scattering of valuable works across various indexing bases, contributing to the dispersion of efforts within the scientific community, as supported by findings in~\cite{fard2019assessing}. From the 788 papers reviewed, 96 papers explicitly defined misinformation and disinformation, while 475 indirectly expressed their understanding of the types of false information, often referring to third-party research. An additional 217 papers provided explicit definitions for at least one type of false information. The confusion in the computer science community's definitions of these terms and their meanings underscores the need for a unified understanding. Furthermore, we found six works dedicated to studying false information taxonomy~\cite{fard2019assessing, lemieux2018leveraging, shu2017fake, kumar2016disinformation, zhou2004building, tudjman2003information}.

\begin{figure}[h]
\centering
\includegraphics[scale=0.5]{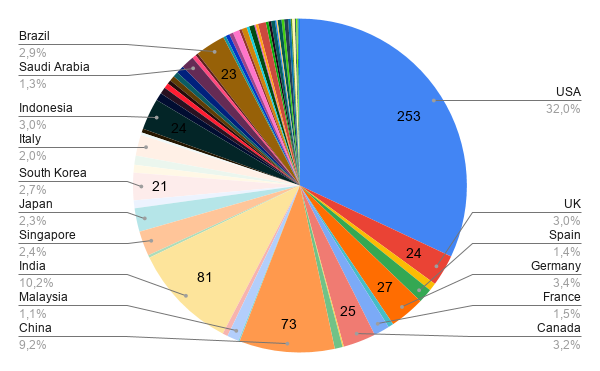}
\caption{Amount of publication by country}
\label{Fig:pub_per_country}
\end{figure}


If the taxonomy challenge is part of the issue, the perspectives of researchers also play a crucial role. Scientific literature suggests a necessity for studies on false information online that consider languages other than English, highlighting the lack of training datasets for automatic detection techniques in non-English languages~\cite{moreno2019factck, pvribavn2019machine, kareem2019pakistani}. Figure~\ref{Fig:pub_per_country} presents distributions by country, symbolizing the subjectivity of perspectives through the cultural diversity of reporting countries, each potentially offering different interpretations of false information terminology and contextual requirements. We delve into a discussion on these aspects, emphasizing the need for sociotechnical research to comprehend social and cultural characteristics that may influence the design and efficiency of solutions.

From a solutions perspective, detection approaches that employed machine learning techniques were predominant~\cite{albahar2019deepfakes, lahlou2019automatic}, considering various criteria for content classification, such as credibility assessment, veracity check, relevancy, fact-checking, bias verification, and trust assessment~\cite{pendyala2019validation}. Solutions on diffusion proposed understanding dissemination patterns and interventions, determining the ideal number and placement of monitor nodes on the network~\cite{pham2019multiple}. Intervention strategies addressed the impact of organizational and technical features in the consumption of false information~\cite{wang2020more}, user behavior regarding uncertainty, susceptibility, trustworthiness, and awareness~\cite{almaliki2019misinformation}, ways of debunking misinformation~\cite{tong2019beyond}, profile characterizations, differences between writing patterns, vulnerabilities~\cite{larson2019applying}, and cyberculture, such as echo chambers~\cite{tang2017echo}.

Regarding false information content, research covered various aspects, including messages with content related to conspiracy theories~\cite{flintham2018falling}, politics~\cite{al2019fake, bedard2018satire}, famous people~\cite{al2019fake}, military, industry, and academia~\cite{sethi2017crowdsourcing, granik2017fake}, financial~\cite{bedard2018satire}, crime~\cite{volkova2017separating}, interesting facts, tips \& tricks~\cite{karadzhov2018we}, hotel reviews~\cite{sandifer2017detection}, product reviews~\cite{yao2017automated}, and health advice~\cite{hailun2014multi}.

\subsection{QP01 - What types of digital false information are studied?}

Our review of 788 articles revealed varied approaches to defining and categorizing misinformation and disinformation. Specifically, 64 studies explicitly delineated their interpretations of \textit{misinformation} and \textit{disinformation}, while 486 articles inferred their conceptualizations through reference to prior works. Additionally, 238 studies explicitly defined one or more specific forms of false information, such as satire and fake news. This analysis underscores a significant challenge in the field: the lack of consistent terminology and definitions for \textit{misinformation} and \textit{disinformation}, complicating the synthesis of literature and hindering effective communication among researchers.

\subsubsection{Diversity in False Information Taxonomy}

\begin{figure}[]
\centering
\includegraphics[scale=0.45]{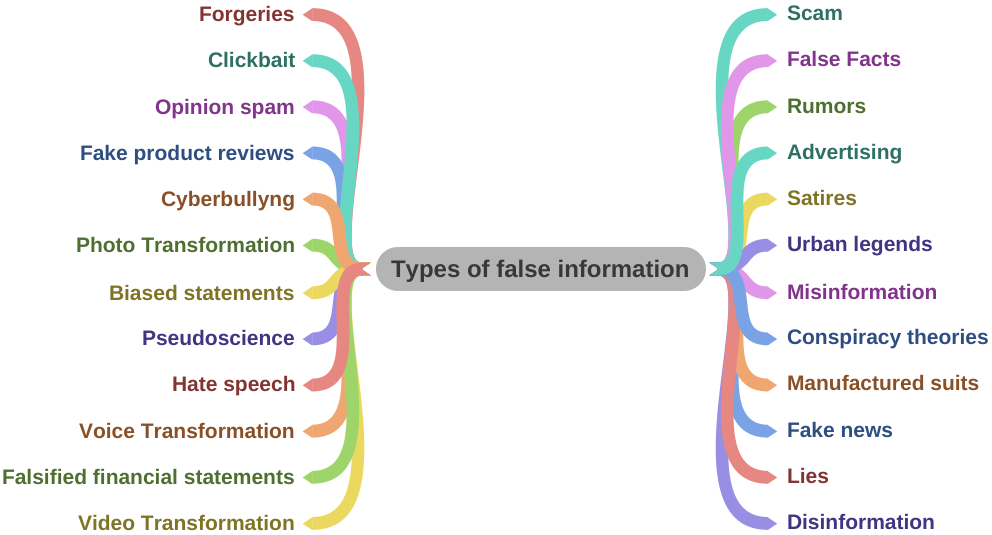}
\caption{Mapping of digital false information types identified in the literature.}
\label{Fig:map-tipos}
\end{figure}

Our investigation into the taxonomy of digital false information uncovered 24 distinct terms employed to categorize digital false information, each with nuanced variations, as illustrated in Figure~\ref{Fig:map-tipos}. For instance, the term "fake news" is variably interpreted across studies, sometimes referring to satirical content \cite{khan2019use}, and other times to hoaxes \cite{ishida2018fake}. This variability and lack of clear definitions challenge the field's progress by obstructing the straightforward comparison and integration of research findings.

Further complicating this landscape, we identified six studies dedicated to the taxonomy and typology of false information, proposing various criteria for classification. In summary, we recognized seven principal criteria across the literature for categorizing digital false information(Table~\ref{tab:criterios}). The predominance of criteria such as evidence, intentionality, and motivation in these studies underscores the complexity of digital false information, necessitating a multifaceted approach for accurate categorization.

\begin{table}[H]
\caption{Criteria for categorization found in the literature.}
\centering
\small
\begin{tabular}{ll}
\hline
Criterion & Description \\\hline
Evidence & whether it is based on evidence or opinions; \\
Significance & if it is about a topic of urgency or perceived importance; \\
About individuals & whether it is about a person or not; \\
Intent & purpose, whether it is intended to deceive or not; \\
Veracity & whether it is true or false; \\
Function & the objective of the message (hurt, explain, manipulate, etc.); \\
Motivation & monetary, personal, political, or other types of gain; \\ \hline
\end{tabular}
\label{tab:criterios}
\end{table}

The most used criteria in the mapped studies are (articles may have considered more than one criterion): evidence (352), intentionality (263), and gain/motivation (41). The categorization of digital false information based on Evidence differentiates it into two types: those based on facts and those based on opinions. These are evaluated based on message quality regarding attributes such as truthfulness~\cite{kumar2013information}, authenticity~\cite{shu2018understanding}, reliability~\cite{fard2019assessing}, and significance~\cite{fard2019assessing}. Regarding the Intentionality, \textbf{dis}information can be accidental (wich also is misinformation) — when stemming from error — or deliberate (disinformation) — constructed/used with the intention to deceive~\cite{shu2018understanding}. The literature reports financial profit~\cite{shu2019beyond}, political gain~\cite{moreno2019factck}, and personal gain~\cite{buchanan2019spreading}.

Table~\ref{tab:terms-results} summarizes the issues with the main types of digital false information.

\begin{table}[t]
\centering
\footnotesize
\caption{Conciseness issues in understanding types of false information}
\label{tab:terms-results}
\renewcommand{\tabularxcolumn}[1]{m{#1}}
\begin{tabularx}{\textwidth}{>{\hsize=0.35\hsize\linewidth=\hsize\centering\arraybackslash}X>{\hsize=0.75\hsize\linewidth=\hsize\raggedright\arraybackslash}X>{\hsize=1.9\hsize\linewidth=\hsize\justifying\arraybackslash}X}
\toprule
Type &
Understanding &
Examples \leavevmode \\ \hline
Rumors &
Ambiguity regarding the Intentionality and Evidence criteria. &
\leavevmode \newline \textbf{Unverified information}\citep{habib2019false, lee2018informed, patel2017modeling, metaxas2015using}; \textbf{That can be true or false}\citep{rana2019rumor, buntain2017automatically}; \textbf{False information}\citep{qin2018predicting, chen2017centralized, zhang2015finding}; \textbf{False propaganda}\citep{tan2019aim}. \leavevmode \\ \hline
Hoaxes &
Conciseness regarding Intentionality. &
\leavevmode \newline \textbf{Deceptive and malicious information} used to deceive and manipulate people\citep{yuliani2018review, hui2018hoaxy}; Intentional \textbf{anti-social content}, such as defamation and bullying~\citep{yu2018adversarial}; \leavevmode \\ \hline
Fake News &
Conciseness regarding the Intentionality and Evidence criteria. &
\leavevmode \newline \textbf{Any form of false information}, from hoaxes to satires~\citep{karduni2018can}; Content \textbf{intentionally created to deceive}\citep{habib2019false, della2018automatic, al2018fake}; Stories posted as \textbf{false facts accepted as genuine}\citep{parikh2018media}; \textbf{Tendentious statements}\citep{murungi2018beyond}; \textbf{Alternative facts} without a basis in reality\citep{purnomo2017keynote}. \textbf{Satires}\citep{chandra2017higher}. 
\leavevmode \\ 
\hline
Satires &
Ambiguity regarding the Intentionality and comedic content criteria. &
\leavevmode \newline Sub-types: parodies, jokes, and pranks\citep{khan2019use, popat2017truth}; \textbf{Accidental}\citep{cybenko2018ai, ishida2018fake}; \textbf{Deliberate false information}\citep{bedard2018satire}; Stories for \textbf{entertainment}~\citep{karduni2018can}. \leavevmode \\ 
\hline
Conspiracy Theory & Ambiguity regarding the Intentionality criterion. Conciseness regarding Evidence. &
\textbf{Intentional fabrications}\cite{tacchini2017some}; \textbf{unreliable information} to explain events or circumstances~\cite{glenski2018humans}; false information, both deliberate and accidental, that simplifies the complexity of social events~\cite{bessi2015viral}; 
\leavevmode \\
\bottomrule
\end{tabularx}
\end{table}

\subsubsection{Discussion}

The persistent challenge within the academic community to reach a consensus on categorizing digital false information typologies is well-documented, spanning from the foundational literature to contemporary debates. The concern over digital false information has been prominent since the advent and widespread adoption of the Internet, as highlighted by \citet{Luciano1996}. Discussions tracing back to 1998 \citep{resnik1998medical}, alongside efforts in the early 2000s to establish a taxonomy \cite{zhou2004building}, underscore the ongoing discord among computer science scholars over defining the multifaceted types of false information \cite{jiang2018linguistic}.

This divergence in definitions, exemplified by the inconsistent categorization of "satire" as either intentional or "accidental misinformation", underscores the necessity for a taxonomy that is not only comprehensive but also sufficiently flexible to encapsulate the complex nature of digital false information. The current categorization criteria, while insightful, often result in ambiguity due to the conflation or overlap of categories such as evidence and veracity, where evidence is purported to encompass veracity as a characteristic.

Moreover, the application of the Function criterion, which encompasses a broad spectrum of intentions (e.g., harm, explain, manipulate), further complicates categorization due to its reliance on the subjective interpretations of those employing the taxonomy. The criterion's effectiveness is contingent upon contextual variables, including cultural nuances and stakeholders' motivations, which can vary significantly across different scenarios.

A notable root cause of the consensus challenge is identified in the foundational literature cited by researchers. For instance, definitions of misinformation draw from diverse disciplines, with \citep{dang2016toward} referencing \cite{rosnow1991inside} to define rumors as unverified statements, while \citep{jin2017multimodal} looks to \cite{allport1947psychology} to describe rumors as deliberate falsehoods or unverified claims. This interdisciplinary borrowing, seen across fields such as sociology \cite{chen2013misinformation} and economics \cite{che2018fake}, contributes to the prevailing ambiguity and hinders a unified understanding of digital false information within computer science.

The current taxonomies' inadequacy in addressing the sociotechnical dimensions of digital false information points to an urgent need for frameworks that more accurately depict the interplay between the technical features of digital false information and its societal impacts. Such an approach would not only aid in demystifying the various types of digital false information but also foster the development of precise mitigation strategies, thereby enhancing the efficacy of interventions aimed at curbing the proliferation of digital false information.

The literature also recognizes the significance of "multimodalities" (e.g., sounds, text, images, videos, gestures) as distinct attributes of false information, underscoring the sociotechnical essence of the phenomenon \cite{jin2017multimodal}. For instance, clickbaits, known for their use of compelling phrases and sensationalist imagery to captivate and monetize audience attention, exemplify how the technical and social attributes of digital false information types can catalyze specific interactions and behaviors.

This examination advocates for a multidisciplinary collaboration to refine the classification of digital false information types. By integrating perspectives from computer science, communication studies, sociology, psychology, and more, the goal is to construct a taxonomy that is not only exhaustive but also flexible enough to adapt to the dynamic nature of digital communication.

\subsubsection{Contribution: A Sociotechnical Framework for Categorizing Types of False Information}

This research synthesizes key findings from the literature to propose a sociotechnical framework for understanding and categorizing the primary types of false information encountered in digital environments. Our framework, presented in Table~\ref{dis_tp}, integrates both the technological aspects of how false information is created and spread, and the social dynamics that influence its reception and believability.

\begin{table}[!h]
\centering
\caption{Sociotechnical Typology of Digital False Information}
\label{dis_tp}
\footnotesize
\begin{tabularx}{\textwidth}{>{\hsize=0.32\hsize}X>{\hsize=0.4\hsize}X>{\hsize=2.28\hsize}X}
\hline
\textbf{Type} & \textbf{Intentionality} & \textbf{Description} \\
\hline
Rumors & Unverified & Messages whose truth value is not initially verified, potentially misleading the audience, often proliferates during crises~\cite{rana2019rumor}. Such messages often lack endorsement from their sources and might cite unverified third parties as authorities, necessitating subsequent fact-checking to ascertain their truthfulness~\cite{devi2018veracity, rana2019rumor}.Their eventual classification as true or false is determined later. \\
\hline
Satire & Deliberate & Crafted to satirically exploit cognitive aspects such as understanding and reasoning, these messages contain comedic elements meant for entertainment~\cite{campan2017fighting}. Though intended to deceive under the guise of humor, audiences are often in on the joke, indicated by implicit cues. Forms include parodies, jokes, and pranks~\cite{khan2019use}. \\
\hline
Hoaxes & Deliberate & Fabricated narratives designed to manipulate the beliefs or behaviors of the target audience. These can range from complex conspiracy theories to sophisticated phishing schemes, leveraging the audience's preconceptions and desires~\cite{yuliani2018review, hui2018hoaxy, goolsby2013cybersecurity, ahmad2019strategically}. \\
\hline
Fake News & Deliberate & A subtype of hoaxes, these false narratives mimic the format of legitimate news to mislead and manipulate public opinion. Unlike satire, the deceptive intent of fake news is not transparent to the audience, often resulting in significant misinformation spread~\cite{wijaya2018improving, manzoor2019fake, murungi2018beyond}. \\
\hline
Conspiracy Theories & Mixed & These narratives simplify complex realities, often attributing outsized influence to malign actors or organizations. They may arise from deliberate misinformation efforts or collective misinterpretations within communities, fostering echo chambers of falsehood~\cite{bessi2015viral, acemoglu2011opinion}. \\
\hline
Clickbaits & Deliberate & Crafted to attract attention and prompt clicks from viewers, these messages often utilize sensationalist or emotionally charged language. The primary motive is typically financial, exploiting user engagement metrics for profit~\cite{glenski2018humans, zannettou2019web}. \\
\hline
\end{tabularx}
\end{table}

Our framework delineates the types of false information by considering both their intentional design (e.g., to deceive, entertain, or profit) and their reception by audiences (e.g., verification challenges, believability). This approach underscores the importance of integrating a sociotechnical perspective in addressing the complexity of digital false information, highlighting the role of human cognition, social norms, and digital affordances in shaping the dissemination and impact of false narratives. Through this classification, we aim to provide a more nuanced understanding of digital false information, facilitating targeted interventions and fostering a critical discourse on the interplay between technology and society in the propagation of digital false information.

\subsection{QP02 - Which stakeholders are considered by the research?}
\label{proto-partes}

Characterizing stakeholders is an important aspect of studying the phenomenon of digital false information, but it has been addressed in a limited manner by the reviewed literature. The research tends to characterize stakeholders in a generic way (136 articles), with classifications such as receiver and sender, interlocutor and audience, producer and consumer, or a reference to the role of "social media user," but with little depth~\cite{resende2018system, traylor2017psyop, hassan2017claimbuster}. In general, there are two main approaches that the literature adopts to address stakeholders.

\begin{itemize}
\item Theoretical models: These models derive stakeholder profiles, characteristics, and roles from cross-disciplinary theories encompassing sociology, psychology, and anthropology. Zhou et al.'s ontological model exemplifies this, delineating stakeholders as contributors, curators, readers, administrators, and analysts, grounded in sociological insights on fraud~\cite{zhou2007ontology}. 
\item Empirical evidence: Empirical studies focus on the human and social dynamics influencing digital false information. They investigate how specific groups discern information credibility across media, contributing to a granular understanding of stakeholder behaviors and perceptions~\cite{piccolo2020pathway, saquete2020fighting}.
\end{itemize}

Regarding the methodology for studying stakeholders, exploratory studies predominated, using interviews and questionnaires that address sociodemographic aspects such as gender, age, and education level~\cite{zhang2018structured, torres2018epistemology, fernandez2018online}, values~\cite{piccolo2020pathway}, sharing motivations~\cite{chin2019new}, among others. Other studies employ ethnographic techniques to analyze behaviors such as sharing actions, verification, and consumption of digital false information in specific groups, such as students and the elderly, in different cultural contexts~\cite{wang2020more, chin2019new, wandoko2019analysis, wason2019building, torres2018epistemology}, or investigate the impacts of homophily and affective closeness~\cite{wu2018toward}. Additionally, some approaches use methods and techniques from Psychology, such as investigations of disinformation consumption patterns based on personality~\cite{8804577}, and differences in the evaluation behavior of objective messages and those with emotional language~\cite{flintham2018falling}. Research that indirectly addresses stakeholders employs complex investigative methodologies, combining methods and techniques, such as~\cite {dang2016toward}, which uses social network analysis, visual analysis, content analysis, and text mining to classify user roles.

\begin{figure}[ht]
\centering
\includegraphics[scale=0.38]{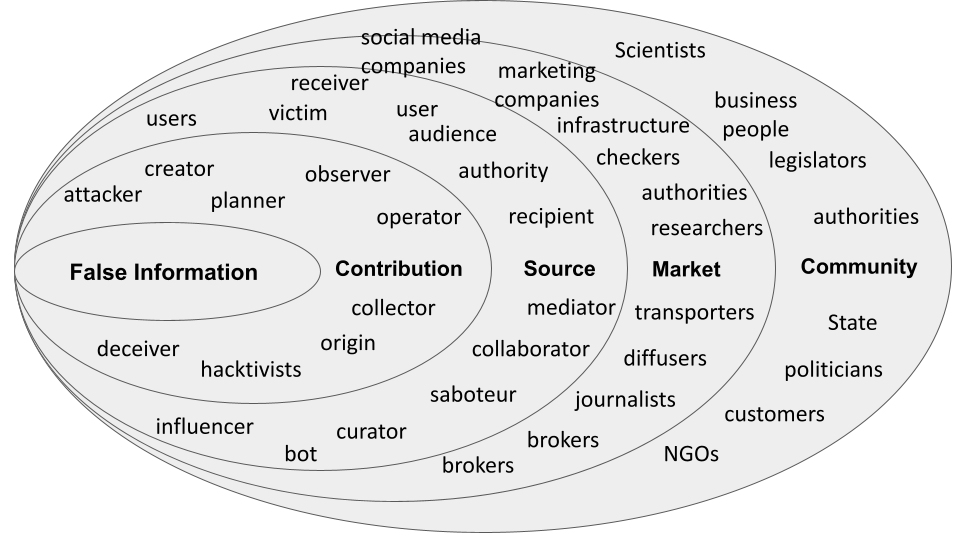}
\caption{Map of stakeholders found in the mapping.}
\label{fig:partes}
\end{figure}

Our synthesis identified 42 distinct stakeholder roles from 236 articles, encompassing a broad spectrum from empirical investigations to theoretical models. We employ the Stakeholder Identification Diagram (SID) for a structured mapping of these roles, encompassing categories like the Contribution group (content creators), the Source group (indirect influencers), the Market (partners and competitors), and the Community (indirectly affected entities) (Figure~\ref{fig:partes}).

\subsubsection{Discussion}

Exploratory methodologies, including interviews and questionnaires, dominate the landscape, targeting demographic and psychographic factors—gender, age, education, values, and sharing motivations. Ethnographic techniques and psychological assessments further enrich our understanding by exploring behaviors and consumption patterns among diverse groups (e.g., students, seniors) and cultural contexts~\cite{wang2020more, fernandez2018online}.

However, these methodological endeavors often fall short in generalizability due to their context-specific findings influenced by cultural norms, beliefs, and limited sample sizes~\cite{lozano2020veracity}. The theoretical models, while providing a structured framework, tend to oversimplify stakeholder roles, lacking in representativeness and failing to account for the intentionality and environmental context of digital false information~\cite{fernandez2018online}.

\subsubsection{Towards an Integrated Framework for Stakeholder Analysis}
To surmount these challenges, we advocate for an integrated framework that combines theoretical insights with empirical evidence, tailored to the multifaceted nature of digital false information. This framework should:

\begin{enumerate}
\item Embrace a multidisciplinary approach, incorporating insights from sociology, psychology, information science, and communication studies to enrich stakeholder analysis.
\item Employ mixed-methods research to balance the depth of qualitative insights with the breadth of quantitative data, enhancing the generalizability of findings.
\item Incorporate adaptive models that recognize the dynamic nature of digital platforms and the evolving tactics of digital false information spread.
\item Highlight the necessity for context-aware analysis that considers the socio-cultural and technological landscapes influencing stakeholder interactions with false information.
\end{enumerate}

This perspective on stakeholder analysis could not only refines our understanding of digital false information dynamics but also could informs the development of targeted interventions and policy recommendations, ultimately contributing to a more informed and resilient digital ecosystem.

\subsection{QP03 - How do the studies address sociotechnical aspects of the phenomenon?}
\label{proto-mdls}

\begin{figure}[h]
\centering
\includegraphics[scale=0.53]{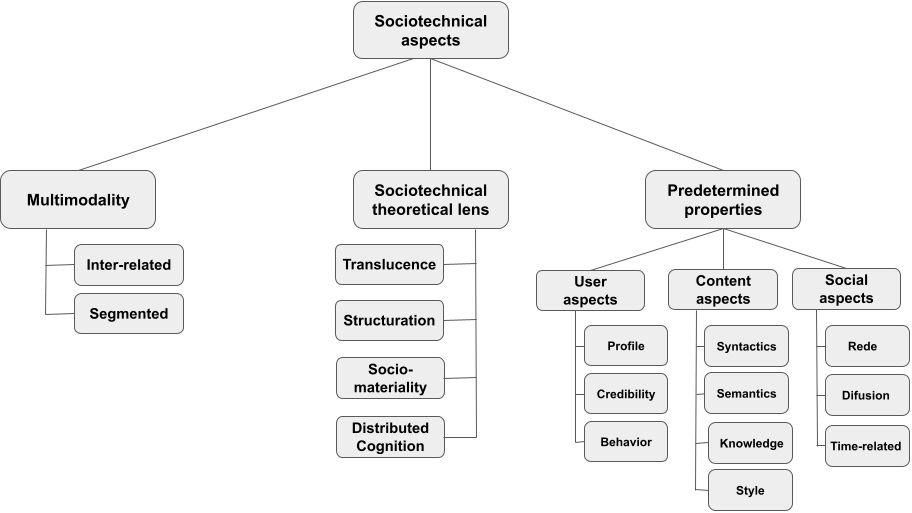}
\caption{Perspectives on approaching sociotechnical aspects}
\label{Fig:sociot}
\end{figure}

The 788 studies were analyzed to identify details about the sociotechnical study of the phenomenon. First, the focus of the analysis was to identify the different ways to approach technical, human and social aspects of the false information phenomenon. We found three different perspectives on approaching sociotechnical aspects (Figure~\ref{Fig:sociot}). Some studies address sociotechnical aspects, classifying them into three levels of abstraction~\cite{zhang2020overview}: user profile aspects, content aspects, and social aspects. Some papers consider the multimodality perspective, addressing communication and meta-communication aspects in different dimensions, such as gestures, sounds, images, text, and layout~\cite{jindal2020newsbag}. Also, some approaches employ theoretical lenses such as Translucence\cite{wang2014think}, Structuration, Socio-materiality, and Distributed Cognition~\cite{starbird2019disinformation}. 

\subsubsection{Predetermined sociotechnical aspects}
\begin{figure}[h]
\centering
\includegraphics[scale=0.5]{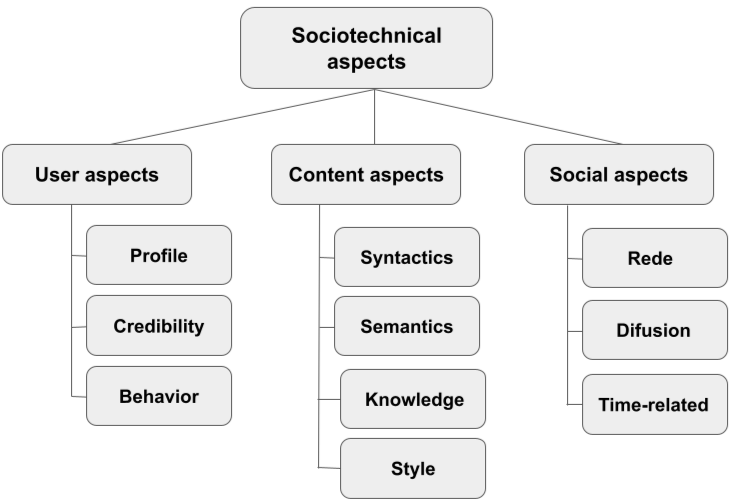}
\caption{Predetermined sociotechnical aspects approach.}
\label{Fig:sociotec}
\end{figure}

Some studies address sociotechnical aspects, classifying them into three levels of abstraction~\cite{zhang2020overview}: user profile aspects, content aspects, and social aspects(Figure~\ref{Fig:sociotec}). These categories have two types of aspects: physical and non-physical. Physical aspects are related to the format and means of communication through which false information spreads. Non-physical aspects are related to human, social, and organizational factors, such as opinions, relationships, and authorities.

User-level aspects are attributes related to the profile of those who create or share false information, and the audience that consumes it. They are subdivided into:
\begin{itemize}
\item Profile attributes: such as name, geolocation information, user registration data (verified or not), whether it has a description or not, and so on~\cite{saquete2020fighting, lozano2020veracity, zhang2020overview}.
\item Credibility aspects: such as the credibility score of the user, number of friends and followers, ratio between friends and followers, total number of posts by the user~\cite{saquete2020fighting, lozano2020veracity, zhang2020overview}.
\item Behavior aspect: such as the user's anomaly score, number of interactions of the user in a time window, the average monthly number of posts by the user, etc~\cite{saquete2020fighting, lozano2020veracity, zhang2020overview}.
\end{itemize}

Content-level aspects are attributes related to the content, format, and aspects of the communication platforms through which false information spreads.
\begin{itemize}
\item Linguistic aspects (Syntactic): syntactic grammatical errors, number of n-grams, term frequency (TF), sentence structure, etc;
\item Semantic aspects: semantic grammatical errors, exaggerated titles, consistency between title and content, contradictions, etc~\cite{saquete2020fighting, lozano2020veracity, zhang2020overview};
\item Knowledge aspects: labeled databases, similarity with verified content, the relationship of the content with the audience's environment, etc~\cite{saquete2020fighting, lozano2020veracity, zhang2020overview};
\item Style aspects: fraction of tweets/posts that contain external links, user mentions, hashtags, popularity of domain names, number of images or videos, clarity score, coherence score, etc~\cite{saquete2020fighting,  zhang2020overview};
\end{itemize}

Social-level aspects are aspects that reflect the diffusion pattern and interaction among users.
\begin{itemize}
\item Network-based resources: clustering of similar users, location, educational background, consumption, and sharing habits~\cite{shelke2019source}.
\item Distribution-based resources: propagation tree, root degree in a tree, the maximum number of subtrees, number of retweets/reposts for an original tweet, a fraction of tweets that are retweeted by an account~\cite{shelke2019source};
\item Temporal resources: interval between posts, post frequency, responses and comments from accounts, time of day when the original information is posted/shared/commented, and the day of the week ~\cite{shelke2019source};
\end{itemize}

\subsubsection{Multimodality}
Thirteen studies on multimodality were found. Multimodality considers the interaction of stakeholders with digital false information as "modes/modalities"\cite{halliday2014language}. Examples of modes include gestures, facial expressions, images, text, layout, etc. Thus, these works address different sociotechnical aspects related to interaction. Predominantly, the studies focus on automatic classifiers that consider \textbf{content-level aspects}, with an emphasis on textual aspects (syntactic and semantic) interconnected with visual aspects (images)\cite{jindal2020newsbag}. Some multimodality studies address the relationship between content-level, user-level, and social-level aspects~\cite{maigrot2018fusion, maigrot2016mediaeval}, considering both technical aspects (such as syntactic features) and social aspects (such as credibility metrics).

\subsubsection{Sociotechnical lens approaches}
Six studies employing sociotechnical lenses were identified. Among them, one study analyzed a sociotechnical system of election\cite{caulfield2019jenny}, another study adopted Translucence sociotechnical lenses to promote responsible behaviors regarding information consumption and sharing~\cite{wang2014think}, and a study on strategic information operations~\cite{starbird2019disinformation} used the theoretical lenses of Structuration, Socio-materiality, and Distributed Cognition to analyze cases of disinformation and identify communication tactics. One study employed the Framing Analysis method to identify biases in news~\cite{7991561} automatically. The other studies were literature reviews of sociotechnical advancements in mitigating the phenomenon.

\subsubsection{Discussions}

As previous literature reviews show~\cite{almaliki2019misinformation, fernandez2018online}, the studies from our literature mapping focus on technical aspects. Even research that considers Social-level aspects integrates content evaluation, topology, and interaction criteria but does not go beyond technical and quantifiable characteristics~\cite{jindal2020newsbag}. For example, \cite{dongo2019credibility} presents a technique for evaluating source credibility considering content, user, and social aspects:

\begin{itemize}
\item Text credibility: based on attributes such as the use of imperatives, ambiguities, and sentimental language;
\item User credibility: based on attributes such as the account creation date and verification status;
\item Social credibility: based on attributes such as followers, shares, and likes.
\end{itemize}

In this example, human and social attributes are technical aspects assigned to individuals related to their actions and characteristics in the virtual environment. However, they do not reflect subjective or informal aspects, such as their ability to perceive and interpret information, values, and beliefs. They are not truly representative of the individual in their context. Thus, these characterizations are helpful only for identifying suspicious activities but do not handle transformations, novelties, and circumstantial factors well. There is a gap between the knowledge that empirical research has been building on sociotechnical aspects and the technical and conceptual productions of interventions and models for the phenomenon. It indicates the difficulty of identifying and interrelating sociotechnical aspects at different levels of abstraction, such as understanding how syntactic and social aspects can influence the consumption of digital false information.

Characterizations of sociotechnical aspects of the phenomenon focus on the relationship between predefined groups of technical, human, and social attributes based on theoretical-methodological lenses from areas such as social network analysis, psychology, and sociology. This approach restricts the flexibility and exploration of other interrelations. An example of such impact is the textual features proposed in the literature, which focus on technical aspects. They are superficial~\cite{zhang2019multi}: number of words, text length, occurrences of "?" and "!" symbols, happy or unhappy emoticons, pronouns in the first, second, or third person, uppercase letters, positive and negative words, Twitter mentions, hashtags, URLs, and retweets. While theoretically grounded, this approach assumes relevant aspects for understanding the phenomenon, wildly generalizing how stakeholders interpret and use these signs in different contexts and for different purposes. This fragmentation of understanding the phenomenon into quantifiable aspects needs to pay attention to relevant qualitative aspects (such as the subjectivity of interpretation by different stakeholders) for these models and characterizations. These characterizations introduce bias in investigating the phenomenon and limit the representativeness of the elements they investigate to develop a solution. It represents a research challenge related to the sociotechnical nature of the phenomenon and an indication of the difficulty of adapting these solutions to deal with the stakeholders' subjectiveness in different contexts~\cite{jindal2020newsbag}.

Solutions that address the interrelation of sociotechnical aspects, such as multimodality, require explicit resources extracted manually from the content and context of the event in which false information occurs~\cite{jindal2020newsbag}. It implies the cost of human resources in analysis and vulnerabilities related to the analyst's bias. Moreover, approaches proposing automatic detection based on multimodal aspects depend on labeled databases or require patterns already identified in other studies for training~\cite{jin2017multimodal}. Thus, the scope of the study of the digital false information phenomenon is limited to the trends and bias presented in the research, methodologies, and attributes, restricting the exploration of new perspectives and characteristics. Additionally, no approaches were found that support disinformation analysis, aid in understanding impact factors, and identify sociotechnical aspects and their interrelationships to characterize false information occurrences in different contexts.

The mapping results show limitations in the approach to interrelated sociotechnical aspects, which, together with the limitations reported in Related Works, indicate the difficulty of dealing with a complex phenomenon like digital false information. Therefore, solutions with integrated strategies and mechanisms that address human, social, and technical aspects at different levels of abstraction are needed to advance the ability to mitigate the harmful effects of the phenomenon.

\subsection{Contribution: semiotic approach for sociotechnical aspects}
\begin{figure}[]
\centering
\includegraphics[scale=0.4]{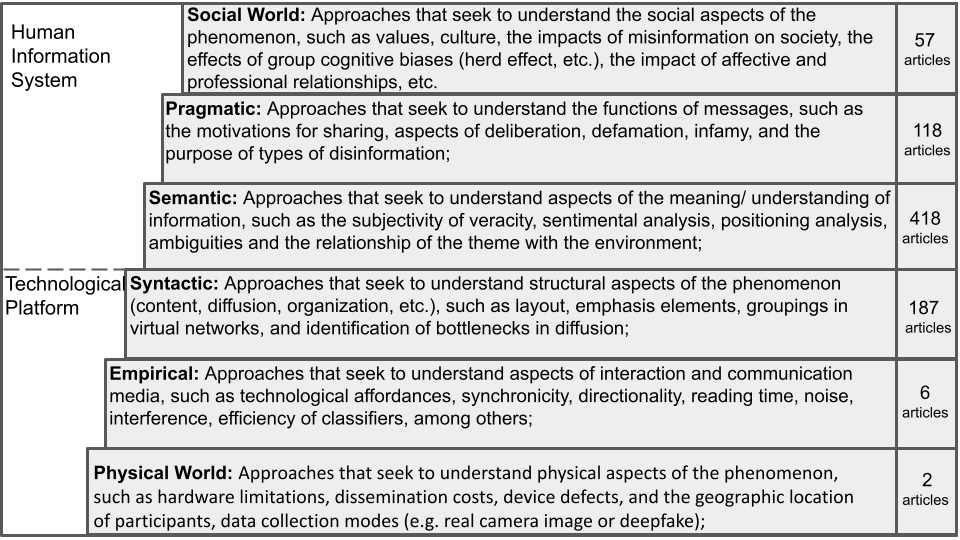}
\caption{Paper categorization scheme.}
\label{fig:es-map}
\end{figure}

The second part of the study was to categorize the studies based on the type of problem and sociotechnical aspects presented in the papers. The articles were analyzed using the content analysis method~\cite{lazar2017research} together with Semiotic Ladder (SL)\footnote{An artifact of Organizational Semiotics used in the analysis of Information Systems~\cite{stamper1994organisational}.}.  The Semiotic Ladder (SL) is a framework used to categorize the properties of an information system into two semiotic dimensions: those related to the Human Information System and those present in the Technological Platform~\cite{liu2000semiotics}. The Human Information System involves the subjective aspects of the perception or conception of a sign, such as the interpretation and construction of meaning. The Human Information System is further divided into Social, Pragmatic, and Semantic levels. On the other hand, the Technological Platform considers how information is formatted, stored, and transmitted, dividing it into Syntactic, Empirical, and Physical levels. We leveraged the SL framework granularity to categorize the studies by examining their coverage of sociotechnical aspects of the digital false information phenomenon and elaborating on their interrelation. The articles are categorized based on the scheme shown in Figure~\ref{fig:es-map}.

The results reveal a greater concentration of studies that address specific aspects at some level of the semiotic ladder with little interrelation. Most sociotechnical aspects are found at the Semantic level (418 articles) and Syntactic level (187 articles). The analyzed literature focuses on sets of technical linguistic features used by classifiers, a tendency to approach the phenomenon fragmented into specific stages without considering the whole, difficulties in addressing aspects of the social level, and limitations in integrating sociotechnical aspects from different levels of abstraction.

There seems to be a limited understanding of the different dimensions of the phenomenon, which has the potential to leave relevant aspects hidden. For example, considering the criterion of Evidence (used in digital false information taxonomies), we observe a tendency in Validation strategies that assess the authenticity of the content of a message based on credibility attributes of the information source, or if other sources corroborate the same content~\cite{lozano2020veracity}. In this regard, there is room for studies and analyses on the Physical and Empirical aspects, such as the artificial production of images, whether an image was captured in the Physical world by some type of sensor (e.g., camera), or generated by a computer (deepfakes), and how they relate to the subjective perception of stakeholders. For reasons of brevity, we grouped the results and discussions into pairs of semiotic dimensions going from the top of the SL to the bottom.

\subsubsection{Social World \& Pragmatics}

We categorized 57 articles addressing aspects of the social world. This category includes research problems associated with group characteristics or group user behavior. There were 26 papers directly involved in investigating social behavior. We categorized eight works on diffusion behavior— seven other researched intervention strategies based on user group awareness and literacy. Table~\ref{Tab:Soc} presents the results.

\begin{table}[ht]
\centering
\caption{Classification of the Social World Dimension}
\footnotesize
\begin{tabular}{llc}
\hline 
Type  & Description & \multicolumn{1}{l}{Qty} \\
\hline 
Behavior Analysis &
   \begin{tabular} [c] {@ {} l @ {}} Researching user consumption patterns in groups, user profiles, \\ based on sociological theories like collective truth and\\ quantitative social metrics like ratings and evaluations. It also\\ works with verification strategies, information correction, user\\ perceptions of credibility, and cognitive capacity. \end{tabular} &
   16 \\
   
False Information Modeling &
   \begin{tabular} [c] {@ {} l @ {}} Research modeling aspects of false information related to the\\ social dimension, such as beliefs, values, relationschips and cultural \\traits. \end{tabular} &
   26 \\
   
Diffusion Dynamics & 
    \begin{tabular} [c] {@ {} l @ {}} Research based on social dimension aspects related to the\\ diffusion behavior of false information, such as sharing\\ practices in groups, audience vulnerabilities \end{tabular} & 08 \\

Intervention Strategies & 
    \begin{tabular} [c] {@ {} l @ {}} Intervention strategies tested in groups, such as increased\\ awareness and digital literacy. 
\end{tabular}                                 & 07 \\   
\hline 
\end{tabular}
\label{Tab:Soc}
\end{table}

Examples in this category include studies on user profile characteristics that make them prone to consuming or dismissing fake news~\citep{shu2018understanding, bay2018social, chen2016bandit, chen2015social}, their impacts on belief formation~\citep{jameel2019social, dimo2019knowing, introne2017collaborative, theng2013dispelling, chen2013misinformation, acemoglu2011opinion}, transformations caused in communication technology~\citep{miletskiy2019transformations}, sociodemographic aspects related to consumption and sharing behavior~\citep{bedard2018satire, chandra2017higher, boshmaf2013design}, in social values like privacy~\citep{cho2016privacy}, in relevant aspects of social media groups, user awareness, credibility perception, and fact-checkers for misinformation diffusion~\citep{wang2020more, zhang2019study, shao2018tracking, murungi2018beyond, safieddine2017spread, hannak2014get}.

We categorized 117 problems in the Pragmatics dimension. This category includes research problems about intentions and their impacts on user consumption, differentiating the roles of misinformation and disinformation in creating false information~\citep{mccarthy2018let}, sharing behavior, and false information campaigns~\citep{wang2018era}. Table~\ref{Tab:Prag} summarizes the categories.

\begin{table}[ht]
\caption{Classification of the Pragmatics Dimension}
\centering
\footnotesize
\begin{tabular}{llc}
\hline 
Type  & Description & \multicolumn{1}{l}{Qty}\\
\hline 
Vulnerability Assessment &
  \begin{tabular}[c]{@{}l@{}}Research investigating user vulnerabilities, such as susceptibility,\\ attack models based on perception manipulation, financial scams.\end{tabular} &
  34 \\
Diffusion Dynamics &
  \begin{tabular}[c]{@{}l@{}}Aspects related to patterns of false information diffusion based on\\ user sharing practices, such as attention propagation.\end{tabular} &
  39 \\
Intervention Strategies &
  \begin{tabular}[c]{@{}l@{}}Intervention strategies related to awareness and literacy tested with\\ users or based on adoption metrics.\end{tabular} &  45 \\
\hline 
\end{tabular}
\label{Tab:Prag}
\end{table}

The pragmatic dimension focuses on research problems related to user behavior in the face of false information. It aims to highlight intentionality, the effects of sociotechnical aspects on user consumption, and the role/purpose of communication/messages. For example, it aims to differentiate between misinformation and disinformation in creating false information~\citep{mccarthy2018let} and examine sharing behavior motivations in false information campaigns~\citep{wang2018era}. It also investigates intervention strategies utility, such as visual cues and verification tools~\citep{nekmat2020nudge, karduni2018can, metaxas2015using, hailun2014multi, wang2014think, chen2015deterring}. Additionally, it explores crowd-based fact-checking~\cite{pinto2019towards}. 

Another category included vulnerability assessment problems, such as manipulation attack models~\citep{raman2019manipulating, shrestha2019online, jansen2012susceptibility, campbell2001and}, exploring cognitive and memory flaws~\citep{raskin2011your, de2009false}, influential node analysis, aspects, influence limitation strategies~\citep{bargar2019challenges, campan2017fighting, liao2012influence, narahari2012influence}, and discussions on legislation for social media responsibility and accountability for orchestrated campaigns.

A research line in the Social World dimension investigates the effects of false information on digital culture. This type of research aids in understanding digital culture and its groups, observing behavior changes, such as how they typically share content, to new information verification tools, benefiting the design of strategies and tools to deal with false information across distinct groups. For example, the transformation of political communication~\cite{miletskiy2019transformations} and changes in how journalism (and journalists) behave online~\cite{starbird2018engage}.

Works investigating the cognitive and psychological characteristics of users affecting false information consumption, such as cues of susceptibility to consuming false information~\citep{jansen2012susceptibility}, the effects of authenticity mechanisms on user behavior~\citep{liao2012influence}, or even motivations for users to share false information~\citep{chen2015social}. For example, research on the manipulation of classification systems that may seem insignificant in long-term time windows~\citep{zhang2019online} but can be affected by social bots and avatars, increasing community consensus in short-term time windows~\citep{wang2018era, yu2018silent, ross2019social}. Additionally, works presenting exploitable cognitive aspects of the user ("user vulnerabilities")~\citep{gunawani2020dentifying} and peculiarities of user beliefs, such as political ideology~\citep{ross2019social}. While understanding user behavior is crucial for developing adequate mitigation strategies, such knowledge has a dark side when employed without due responsibility. It can be used with malicious intent for manipulation in targeted marketing campaigns~\citep{ahmad2019strategically, beskow2019agent, miletskiy2019transformations, bevensee2018alt, bandeli2018analyzing}.

Regarding responsibility, the research results in the Pragmatic dimension point to different uses for digital false information, such as fake news and deepfakes, like sophisticated social engineering attacks~\cite{gunawani2020dentifying, fraga2020fake}. Discussions on orchestrated digital false information campaigns~\citep{bandeli2018analyzing} and information warfare~\citep{nestoras2018political, loui2017information} designed to deceive the user based on semantic and cognitive hacks~\citep{cybenko2002cognitive, thompson2003semantic} highlight the need for legal responsibility. \citet{watney2018legal} discusses the legal dimension of fake news, the legal position of social media, and the need for state regulation. It deliberates on the inadequacy of self-regulation measures by social media and the ongoing efforts of governments legislating to define the responsibilities of information guardians.

\subsubsection{Discussion of the Social World \& Pragmatic Dimensions}

Research on the Social World dimension considers cultural and social aspects of the digital false information phenomenon. Understanding the intricacies of group interaction, from modeling the reasoning process~\citep{wang2020more} to the effects on belief systems~\citep{dimo2019knowing, jameel2019social}, is crucial for broadening the understanding of digital false information as a phenomenon. The intentionality and goals of individuals involved in a disinformative event, be it the audience or the creators, impact communication. First, understanding the deceptive intentions behind digital false information is challenging. It may include social and benevolent reasons, such as lying about a surprise party or showcasing belonging to a community~\cite{karlova2013social}. It may have antagonistic personal motives, like selling a broken device online or ruining someone's reputation~\cite{karlova2013social}. Furthermore, deliberate digital false information can be created purely for amusement~\cite{mccarthy2018let}. In this sense, characterizing the stakeholders constitutes the essence of digital false information communication, forming part of the information system structure that contextualizes a digital false information occurrence.

Although critical, research in the Social World dimension also raises concerns about responsibility. To what extent are tools made to assist users helpful? Some solutions promote potentially harmful and persuasive technology. For example, mechanisms that can induce changes in consumption behavior through nudges~\citep{nekmat2020nudge, bhuiyan2018feedreflect}, warnings, and automated technology to mitigate false information~\citep{goindani2020social, ookita2017effective, chen2015deterring, wang2014think}. Human-computer interaction, especially intermittent positive reinforcement mechanisms, has the potential to harm users, increase user anxiety, and disrupt the user's life in other dimensions~\citep{10.1145/3204447, 10.1145/2591708.2591756, bailey2001effects}. For instance, a feedback-based solution that motivates users to share the truth, estimating the responses a particular post might receive~\citep{goindani2020social}. One can speculate the scenario of user self-censorship. Invasive tools that assist the user must undergo careful development. To what extent does constructing "persuasive solutions" improve the user experience? More discussion on ethical design and implementation of such measures is crucial to responsibly building a better digital social world.

Education also plays a role in the Social World and Pragmatic research as a weapon to combat online digital false information, enhancing user information consumption behavior. Teaching ways to improve critical thinking in distinct user groups, such as classrooms or social media users~\citep{tanaka2013toward, pollalis2018classroom}, is essential for developing the user's ability to detect and engage in fact-checking activities and promoting responsible behavior.

\subsubsection{Semantic \& Syntactic}

We found 418 articles on the Semantic dimension. This dimension includes research on meaning, bias, and sentiment analysis for fake information detection. Table ~\ref{Tab:Sem} summarizes the categories.

\begin{table}[ht]
\centering
\caption{Classification of the Semantic dimension}
\centering
\footnotesize
\begin{tabular}{llc}
\hline 
Type & Description & \multicolumn{1}{l}{Qty} \\
\hline 
Fake information detection & \begin{tabular}[c]{@{}l@{}}Research dedicated to characterizing and identifying false\\ claims using machine learning.\end{tabular} & 217 \\
Literature reviews & \begin{tabular}[c]{@{}l@{}}Reviews related to detection models and information\\ quality assessment techniques.\end{tabular} & 39 \\
Automated fact-checking & \begin{tabular}[c]{@{}l@{}}Research dedicated to verifying the truthfulness\\ of claims based on semantic criteria like\\ semantic proximity between texts and use of\\ sentimental language.\end{tabular} & 106 \\
False information models & \begin{tabular}[c]{@{}l@{}}Research characterizing a type of false information,\\ such as semantic aspects affecting consumption.\\ Examples include political biases, rhetorical discourse, and\\ logical fallacies.\end{tabular} & 15 \\
Diffusion models & \begin{tabular}[c]{@{}l@{}}Research characterizing diffusion patterns.\end{tabular} & 21 \\
Datasets & \begin{tabular}[c]{@{}l@{}}Labeled datasets, debunked fake news, and multimedia\\ used for benchmark purposes.\end{tabular} & 20 \\
\hline 
\end{tabular}
\label{Tab:Sem}
\end{table}

Examples include semantic feature analysis for characterizing fake information~\citep{vereshchaka2020analyzing, oehmichen2019not, zhou2004building}, credibility assessment~\citep{hassan2018text}, truthfulness evaluation~\citep{lal2018check}, bias assessment~\citep{patankar2019bias}, and~\citep{yasser2018bigir}. Research on diffusion patterns considering content analysis~\citep{budak2019happened, zhang2019study}, sentiment analysis~\citep{del2017news, dang2016toward}, and semantic analysis~\citep{huang2019dependable, broniatowski2019illuminate, tschiatschek2018fake}. All literature reviews are in this category.

We found 177 works in the "Syntactic" dimension. This group includes works related to formal structures of diffusion and message content. Ninety-six articles on automated detection strategies focused on image alteration or syntax-level analysis. We found 65 works on dynamics and diffusion patterns. There are 25 evaluations of intervention strategies and tools. Table ~\ref{Tab:Sin} summarizes the categories.

\begin{table}[ht]
\caption{Classification of the Syntactic dimension}
\centering
\footnotesize
\begin{tabular}{llc}
\hline 
Type & Description & \multicolumn{1}{l}{Qty} \\
\hline 
Fake information detection & \begin{tabular}[c]{@{}l@{}}Models for fake information detection, early detection,\\ syntax-based methods like Levenshtein distance and\\ multidimensional vectors, and source credibility assessment.\end{tabular} & 96 \\
Diffusion dynamics & \begin{tabular}[c]{@{}l@{}}Research on dynamics and control of diffusion,\\ structural aspects of the social network, where to cut links,\\ which nodes to intervene, characteristics of competing campaigns.\end{tabular} & 65 \\
Intervention strategies & \begin{tabular}[c]{@{}l@{}}Research on usability and characterization of intervention\\ strategies and tools, like fact-checking website features,\\ weakness of intervention strategies, profit minimization, nudges.\end{tabular} & 26 \\
\hline 
\end{tabular}
\label{Tab:Sin}
\end{table}

In this group, we found works conducted at the structural level of messages, such as characterization of fake information~\citep{karimi2018influence, rehm2017infrastructure}, investigating criteria for source credibility assessment~\citep{pacheco2019yakuin, fernandez2011healthtrust}, and fake account detection~\citep{santia2019detecting, khaled2018detecting, kumar2014accurately}, audio/video manipulations~\citep{huh2018fighting, chen2018focus}, fake information databases~\citep{rubin2012art, kareem2019pakistani, kapusta2020improvement}, and infrastructure models~\citep{rehm2017infrastructure}.

\subsubsection{Integration of Semantic \& Syntactic Dimensions}

Numerous studies have explored the characteristics of fake information by analyzing the relationship between its Semantic and Syntactic dimensions. Thirty-nine works model various types of false information. Ontology-based approaches have been used to identify attributes of digital false information, such as type, motivation, origin/destination, communication channel, date/time of onset, evidence, and confidence~\cite{zhou2007ontology}. Some researchers have also investigated properties of fake information on web content, such as relevance~\citep{lin2009identifying}, anxiety and informational certainty~\citep{oh2010exploration}, trust~\citep{mendoza2010twitter, cholvy2014strong}, deception~\citep{hussain2018analyzing, traylor2017psyop, rubin2015deception}, and diffusion behaviors and patterns~\citep{broniatowski2019illuminate, kumar2020anatomical}.

In detection strategies, we categorize works into two main paths: early detection and late verification. Early detection methods are designed to identify false information as quickly as possible to prevent its spread to a large part of the social network or to prevent infection of other groups~\citep{kumar2020anatomical, budak2019happened, del2017news}. They consider multidimensional aspects, such as relevance assessment~\citep{yasser2018bigir}, diffusion patterns~\citep{kumar2020anatomical, budak2019happened}, and the influence score of social network nodes~\citep{yang2019crisis}. Examples of mitigation strategies include cutting propagation links~\citep{ruan2015efficient}, containing diffusion in influential nodes~\citep{yang2019crisis}, and improving positive information cascades~\citep{farajtabar2017fake}. The latest verification methods are built to check information that is already spreading. They rely on the similarity of aspects of older identified fake information to detect the spread of new fake information~\cite{sethi2018extinguishing, pourghomi2017stop, leblay2017exploring}. Both consider multidimensional aspects, such as content emotion~\citep{sethi2018extinguishing}, stance classification~\citep{xuan2019rumor}, and credibility of profile and content~\citep{nilforoshan2019slicendice}.

We also found 20 datasets on fake information content. The diversity of languages and purposes ranges from Czech~\citep{pvribavn2019machine}, Polish~\citep{pvribavn2019machine}, Slovak~\citep{pvribavn2019machine, kapusta2020improvement}, Pakistani~\citep{kareem2019pakistani}, Portuguese (Brazil)~\citep{silva2020towards, moreno2019factck}, and Arabic~\citep{alkhair2019arabic} to Multimedia~\citep{kopev2019detecting}, and Image Forgery Detection Dataset~\citep{rahman2019smifd}. There were 12 other datasets in the English language.

\subsubsection{Discussion of Semantic \& Syntactic Dimensions}

Research in the Semantic and Syntactic dimensions investigates aspects related to the content characteristics of fake information, the virtual organization structures of user groups, diffusion patterns, and influential aspects for understanding and truth assessment. Characterizations, often derived from automated detection methods, have limitations related to the multidisciplinary complexity of the aspects involved in the occurrence of digital false information in communication. In this sense, the studies seek to determine an appropriate set of features capable of indicating false information to improve classifier accuracy~\cite{katsaros2019machine}. The most prominent methods consider multidimensional criteria, analyzing user profile features such as whether a profile is verified, diffusion characteristics such as social connections or the number of shares in a time window, and content aspects such as emotional appeal~\citep{cordeiro2019real}. However, these pre-determined sets of aspects limit understanding of multidisciplinary complexity and incredibly informal aspects of communication. Values, culture, and beliefs are informal aspects hidden from these approaches, reflecting the tendency to address technical aspects that characterize a profile, content, or diffusion but do not consider relationships with the audience environment. In this sense, automatic classifiers are underutilized and can potentially cause harmful consequences, such as filter bubble alienation and ideological echo chambers~\cite{kumar2020anatomical}.

Another consideration regarding automated detection methods is their adaptability to alternative forms of false content. If disinformation is intentionally created to manipulate, it is reasonable to assume that the designers' disinformation techniques were developed to avoid known detection strategies~\citep{gray2020Nov}. Since automatic detection uses fake information models to analyze false content, some limitations exist regarding contextual characteristics beyond performance and accuracy canons~\cite {zhou2019fake}. Dealing with human communications requires understanding human behavior and informal nuances, such as culture, values, and beliefs, shaping how each community thinks and communicates. Additionally, simple classification into the true or false dichotomy (fact-checkers and automated strategies) fails when facts alone do not convince people, but the persuasive appeal rhetoric linked to the belief structures of a particular social group does~\cite{murungi2018beyond}.

There is a need for datasets in languages other than English. The results indicate that English datasets are the most prominent, with 12 articles reporting them. The effectiveness of mitigation strategies depends on contextual information and false resources, such as content, user, social, and network. The diversity of features in the Syntactic and Semantic dimensions of fake information models in alternative languages enhances the benchmark evaluations of automated classification methods. Furthermore, the nuances of fake information phenomena in the "Pragmatic" and "Social World" dimensions of different communities extend the actions of mitigation strategies and tools, introducing alternative perspectives to address the issues.

\subsubsection{Empirical \& Physical World}
There were eight works in the Empirical and Physical World dimension. Table~\ref{Tab:emp} summarizes the categories.

\begin{table}[ht]
\centering
\caption{Classification of Empirical and Physical World dimensions}
\footnotesize
\begin{tabular}{llc}
\hline 
                      Type  & Description & \multicolumn{1}{l}{Qty}\\
\hline 
Comparative detection models &
  \begin{tabular}[c]{@{}l@{}}Research exclusively dedicated to comparing the\\ performance of automatic fake information detection models.\end{tabular} &
  6 \\
Structural properties of diffusion &
  \begin{tabular}[c]{@{}l@{}}Studies on the economy of fake information diffusion,\\ such as sponsors, and the infrastructure for diffusion\\ mitigation.\end{tabular} &  2 \\
\hline 
\end{tabular}
\label{Tab:emp}
\end{table}

In the Empirical category, we discuss issues related to the structure of signs and the interaction features of digital objects and technologies (digital and technological affordances). It includes evaluating the performance of state-of-the-art detection algorithms like the one presented in ~\citet{jeong2019learning} and heartbeat-based detection of fake multimedia videos ~\cite{fernandes2019predicting}. Some works focus on logical connections and software constraints for designing and diffusing fake information~\cite{zhu2018diffusion, elkasrawi2016you, ruan2015efficient}. Lastly, this category covers research on cryptography challenges ~\cite{tyagi2019traceback, melo2019whatsapp}. On the other hand, the Physical World category looks at the rawest level of an information system device and methods for obtaining information. It includes studies on image compression aspects based on detection~\citep{nikoukhah2018automatic} and studies that consider the different aspects of an image collected from a digital camera and one created digitally (e.g., deepfakes).

\subsubsection{Discussion of Empirical and Physical World dimensions}
For the Empirical and Physical World dimensions, there are considerations regarding access to private content on personal communication apps, such as WhatsApp~\citep{de2019can, tyagi2019traceback, melo2019whatsapp}. This type of communication software differs from an open communication platform, where user-generated content has customized privacy settings. Messages in these apps are private, and accessing them constitutes a privacy invasion, even if done by a machine. It is in the interest of the diffusion structure administrator to mitigate any content that may harm users. For instance, Facebook and Twitter regulate content on their social networks to prevent harmful health advice. WhatsApp has different requirements to deal with fake information issues, such as the privacy monitor exchange. Additionally, deceptive information phenomena may affect and impact network service providers.

Studies on diffusion and content creation provide a perspective on the technical functions that affect the fake information phenomenon. For example, encryption makes detection and monitoring of spread a challenging task~\cite{tyagi2019traceback, melo2019whatsapp}. Furthermore, hardware and software properties affect the user's communication process. For instance, a faulty network connection can impact received information and how the user consumes it, introducing bias. Understanding the constraints and enhancements of technological properties in consumption and diffusion processes is necessary.

\subsection{Contribution: interplay between sociotechnical aspects}

The Semiotic Ladder (SL) provides a segmented view of an organization or information system. Building upon this perspective, we can categorize the sociotechnical aspects of digital false information at each semiotic level of SL. The results guide the discussion of the interplay between sociotechnical aspects in instances of digital false information in communication, indicating possible avenues to advance understanding the phenomenon. We can observe the importance of comprehending digital false information comprehensively—as a sociotechnical phenomenon systemically. 

The Social World dimension reveals relevant social context factors in the occurrence of digitally false information in communication. Since sociodemographic aspects such as economic inequality, digital literacy, or internet access are structural components of the phenomenon, understanding the contextual aspects leading to digital false information in communication is crucial. These aspects represent limits to communication technologies and access to information, reflecting how a message is interpreted/represented. Informal contextual aspects, such as values, expectations, and beliefs, help understand the framing that a message employs to communicate an idea and the framing used by the audience to comprehend it. In this sense, digital false information can be accidental—when there is no intention to deceive, but a deviation occurs from the intended interpretation by the stakeholder delivering the message—or deliberate—when the intention is to manipulate the interpretation of a particular situation/message.

It is important to understand the connection between the Social World aspect and Pragmatics to distinguish between different types of false information. While both types of digital false information may present factual information, they can be designed to introduce biases and influence the user's perception and understanding of a situation through narratives and framing techniques. Several studies, such as ~\citet{aigner2017manipulating, karlova2013social}, have provided evidence of how social media can be manipulated to spread digital false information. Some propaganda strategies appeal to cognitive biases triggering social behaviors~\cite{volkova2017separating, mccombs1972agenda}, indicating the intentionality of communication. Other examples explore cognitive dissonance~\citep{larson2019applying, bai2019exploring}, the use of persuasion weapons~\citep{varol2018deception}, semantic social engineering~\citep{java2019detection}, cognitive hacks~\citep{cybenko2002cognitive, java2019detection, bai2019exploring}, and semantic hacks~\citep{thompson2003semantic}. In this perspective, deliberately false information (disinformation) can be considered part of a coordinated effort to deceive, persuade, and manipulate. Thus, malicious intent is implicit in those who deliberately create and disseminate false information~\citep{da2019fake}. Pragmatics is a crucial dimension in the discussion of legal accountability, from self-regulation measures of social media to ongoing government efforts legislating to define the responsibilities of digital false information carriers.

The relationship between the Social World, Pragmatics, Semantics, and Syntax dimensions—the structural aspects of digital false information messages—is significant. For example, the characteristics of plausibility and believability of false content related to facts and ongoing events can increase the credibility of the content topic~\citep{8804577, rath2017retweet} and manipulate users' perception of credibility~\citep{aigner2017manipulating}. In this sense, the relationship between the theme and the audience's environment must be considered when analyzing a digital false information occurrence. \citet{murungi2018beyond} asserts that the danger of deliberate digital false information may not lie in its deviation from the fact but in its persuasive appeal according to the pre-existing beliefs of a particular social group. When studying the spread of digital false information, it is essential to consider the informal social environment in which it occurs. It is especially true when analyzing how people interpret digital false information as truth and the intentions behind consciously sharing lies. The semantic and pragmatic aspects of digital false information should be considered to gain a comprehensive understanding of the phenomenon.

The technologies used for message transmission are at the physical and empirical levels. These dimensions, interrelated with the Social World, Pragmatics, Semantics, and Syntatics, reveal various impactful aspects. For example, hyperconnectivity induced by pervasive mobile computing allows people to receive false information anytime and anywhere~\citep{berghel2017software}, especially during significant events such as terrorist attacks, accidents, and natural disasters~\cite{indu2019nature}. 
Empirical aspects of technology shape the communication process, limiting the speed and range of diffusion. Mobile communication or broadband internet access, the form of transmission, whether unidirectional or bidirectional, simultaneous or not, impacts the dynamics of information communication and audience interactions~\cite{Luciano1996}. For example, a journalistic portal or a web page functions like radio and TV, communicating only to its audience that can receive information. On the other hand, WhatsApp allows the exchange of information between the interlocutor and the audience, enabling challenges and direct questions to clarify dubious points in the discourse. Encryption is also relevant, considering digitally exchanged messages can be encrypted~\cite{tyagi2019traceback}. Cryptography complicates automatic detection or diffusion control tools, and breaking it may characterize an invasion of privacy~\citep{puska2020whatsapp}.

The way digital social network organizations occur, like an opportunistic or static ad-hoc mobile network~\cite{wang2008defending}, imposes limits on digital false information diffusion and consumption. For example, WhatsApp uses the mobile service infrastructure to interconnect nodes known by their smartphones or an opportunistic communication app that builds a social network when nodes are physically close~\cite{tyagi2019traceback}. Another aspect is the use of automated digital resources, such as micro-targeted marketing campaigns for opinion modeling~\citep{zhang2020overview, lee2019detection}, and the use of bots for sharing~\citep{bastos2019brexit}, which play a role in the reach and speed of digital false information dissemination. In this sense, digital affordances contribute to the diffusion and consumption of digital false information~\cite {basu2019identification}. Moreover, information cascades—information spreading in cascading diffusion patterns through organic sharing~\cite{jameel2019social}—are linked to informal aspects of the social and human context\cite{10.1145/3320435.3320456}, and the technological affordances of artifacts and communication media\cite{starbird2019disinformation}.

Moreover, economic aspects are another critical variable in the digital false information phenomenon~\cite{kshetri2017economics}. The characteristics of the physical device used for the communication process and how people interact with it. For instance, the theory of long-term effects in communication science~\citep{wolf2003teorias} advocates for the gradual construction strategy of meanings of some content consumed by the public, flooding a propaganda message in different forms and media. Similarly, the user may receive the same disinformative content on different applications on the same or different devices they own~\cite{bandeli2018analyzing}. Restrictions on access to journalistic or corrective information hinder verification and mitigation activities of digital false information~\citep{kshetri2017economics}.

On the other hand, users may be stuck with restrictive access plans, low-cost devices that make their usage experience slow, tedious, and tiresome, and various applications with their own rules for blocking and distributing content. Furthermore, influential individuals with thousands of followers disseminate digital false information received/discussed in other media on their profiles, monetizing the views their profile receives. The physical dimension encompasses crucial aspects to understand better the phenomena and the blind spots where technology and design fail.

Social media gather information on individuals' profiles and behaviors and sell them to third parties for marketing purposes~\cite{bastos2019brexit}. Additionally, automated methods for collecting information from online social networks, such as web crawlers, can be used to build user profiles and identify groups of interest~\cite{7321230}. Even technical interface features and interaction resources sometimes reveal more than necessary, exposing critical user information to third parties~\cite{puska2020whatsapp}. \citet{murungi2018beyond} attests that the danger of deliberate digital false information may not lie in its deviation from the "fact" but in its persuasive appeal according to the pre-existing beliefs of a particular social group. Thus, users' personal and behavioral data constitute another contextual aspect to consider in analyzing digital false information events, as they impact persuasion, influence, and human manipulation.

In this sense, it is necessary to support the study concerning the interrelation of sociotechnical aspects and their impacts on dis/misinformation cases.

\subsection{Contribution: mitigation strategies categorization}
\label{Appendix-Strategies}
Mitigation involves activities related to combating digital false information and interventions to reduce its effects. Mitigation strategies can be executed automatically, in a hybrid manner, or entirely manually by stakeholders~\citep{parikh2018media}. Automatic mitigation involves technical artifacts that mechanize control and accountability actions, such as automatic detection mechanisms or labels that mark a specific piece of digital false information. Manual mitigation includes strategies that support stakeholders in the task of identifying digital false information, such as solutions integrated into the browser that facilitate message verification~\cite{pourghomi2017stop}. Hybrid strategies combine technical aspects and functions of human information systems to enhance the efficiency and effectiveness of detection and corrections, such as the automatic detection of suspicious messages to be verified by human agents~\cite{volkova2017separating}.

We did not find a categorization for digital false information mitigation strategies. In this regard, this literature review understands that deliberate digital false information mitigation is an activity related to information security. Therefore, digital false information mitigation strategies were characterized based on the information security countermeasures taxonomy~\cite{avizienis2004basic}.

Mitigation activities are grouped into four categories of countermeasures: Prevention, Prediction, Removal, and Tolerance. Prevention strategies aim to prevent the "contamination" of stakeholders by known disinformative objects, preventing their consumption and creation and blocking their initial dissemination. Prediction strategies intend to estimate the existence of vulnerabilities that can be exploited through deliberate digital false information, such as identifying trends, evaluating limitations in operation protocols, and susceptibility analyses~\cite{wilder2018controlling}. Moreover, these aspects also indicate weaknesses that can lead to accidental occurrences of digital false information. Removal strategies consist of spreading corrections, correcting known vulnerabilities in an organization, such as updating detection algorithms and training stakeholders. Tolerance strategies aim for the organization's resilience in the presence of digital false information, such as the use of automatic detection mechanisms, and awareness and training campaigns. Table~\ref{tab:mitigation-strategies} provides examples of mitigation strategies and activities in each category.

\begin{table}[!h]
\caption{Digital False Information Mitigation Strategies.}
\footnotesize
\label{tab:mitigation-strategies}
\renewcommand{\tabularxcolumn}[1]{m{#1}}
\begin{tabularx}{\textwidth}{>{\hsize=0.3\hsize\linewidth=\hsize\raggedright\arraybackslash}X>{\hsize=0.9\hsize\linewidth=\hsize\raggedright\arraybackslash}X>{\hsize=0.8\hsize\linewidth=\hsize\raggedright\arraybackslash}X>{\hsize=1\hsize\linewidth=\hsize\raggedright\arraybackslash}X}
\hline
Type & Description & Activities & Examples\\ \hline
Prevention &
Avoid the contamination of the organization by known disinformative objects, preventing their consumption and creation, and blocking their initial dissemination. &
Detection; Blocking; Training; Awareness; Standardization; &
Spam detection; phishing fraud detection; social engineering training; awareness of consequences; Legislation to combat fraud, defamation, etc. \\ \hline
Prediction &
Intend to estimate the existence of vulnerabilities that can be exploited through deliberate digital false information in an organization and limitations that may lead to digital false information in communication. &
Risk assessment; Vulnerability identification; Forensic analysis; &
Behavior modeling, susceptibility assessment, reliability and credibility assessment; \\ \hline
Removal &
Consist of correcting digital false information and mitigating known vulnerabilities that may lead to digital false information in communication. &
Maintenance; Training; Awareness; &
Update automatic detectors, removal of disinformative objects, disinformative object modeling, pattern identification in interactions, account censorship; \\ \hline
Tolerance &
Aim for the organization's resilience in the presence of digital false information. &
Detection, Blocking, Decontamination; Correction; Training; Standardization &
Diffusion monitoring, content pattern identification, diffusion control, training and awareness, etc.; \\ \hline
\end{tabularx}
\end{table}

Detection mitigation includes the collection of disinformative objects~\cite{dang2016toward}, pattern recognition\cite{wai2018identifying}, verification of veracity~\cite{devi2018veracity}, use of digital tools to aid in the identification of disinformative objects, such as visual annotations in messages~\cite{gao2018label, wood2018rethinking}, and manual mechanisms for detection\cite{pourghomi2017stop}, as well as automatic tools for detecting disinformative objects and accounts~\cite{wilder2018controlling}. Blocking includes cutting links~\cite{ruan2015efficient}, blocking accounts~\cite{khaled2018detecting}, removal of disinformative objects~\cite{krishnamurthy2013mis}, diffusion control, sharing limitations~\cite{melo2019whatsapp}, and more.

Training and awareness activities include the promotion of responsible behaviors~\cite{wang2014think}, content verification~\cite{Torres2018epstem}, enhancing stakeholders' risk perception~\cite{almaliki2019misinformation, caputo2013going}, and teaching verification techniques~\cite{ireland2018fake}. Risk assessment activities consider susceptibility discovery~\cite{jansen2012susceptibility}, the inference of content credibility and reliability and its sources~\cite{dongo2019credibility, cota2019quantifying}, the prediction of future rumors~\cite{qin2018predicting}, simulations of digital false information events~\cite{bossetta2018simulated}, and communication models of disinformation~\cite{traylor2017psyop, zhou2007ontology}, among others. Mitigation also deals with formal activities such as legislation, standards, and statutes. False information in cyberspace requires new laws~\cite{rowe2004two, rowe2006taxonomy}, and legislative instruments for control and accountability~\cite{watney2018legal}.

Maintenance includes the creation and updating of databases~\cite{rubin2012art}, the improvement of automatic detection mechanisms~\cite{kapusta2020improvement, poddar2019comparison}, modeling of disinformative objects~\cite{cho2017modeling, mack2007models}, and the identification of sharing motivations~\cite{wason2019building, pal2018salient, chen2013misinformation, zhou2007ontology}.

Corrections of digital false information can be carried out by any interested party, from specialized organizations such as fact-checkers to amateur groups and even bots on social networks. For the effectiveness of this type of mitigation, it is important that corrections are disseminated through the communication channels that users are accustomed to using~\cite{fernandez2018online}. In addition to corrections, the disclosure of the motivations, consequences, and origins of digital false information influences stakeholders' consumption of corrections~\cite{fernandez2018online}. Another important aspect of this type of mitigation is the relevance of the stakeholder (authority, influencers, experts, etc.) who disseminates the corrections to the group~\cite{fernandez2018online}.

\subsection{QP04 - How do research approaches address the organization of digital false information case?}

Generally, there are two ways to approach the phenomenon~\cite{zhang2019multi}: post-level approaches --- which deeply examine a single social media post --- and event/case-level approaches --- which analyze a group of posts that constitute an event/case. Regarding the study of communication, whether at the post or event level, research on the phenomenon presents four main perspectives to address communication:

\begin{itemize}[noitemsep]
\item Non-systemic approaches: These studies focus on particular stages of a piece of digital false information's life, such as detection studies in diffusion~\cite{pourghomi2017stop}, news consumption behavior~\cite{flintham2018falling}, and processes of creating fake news~\cite{mccarthy2018let}.
\item Systemic approaches to disinformation diffusion: These studies examine the life of disinformation considering interconnected stages, focusing on the perspective of those who construct the disinformation, such as stages of disinformation campaigns and deception strategies~\cite{starbird2019disinformation, traylor2017psyop}.
\item Systemic approaches to diffusion: These studies consider interconnected stages from the perspective of diffusion, such as the emergence-diffusion-resurgence model~\cite{shin2018diffusion}, emergence-diffusion-decay model~\cite{chua2016collective}, ignorant-diffuser-suppressor model~\cite{nekovee2007theory}, or the susceptible-infected-recovered epidemic model and variations~\cite{cho2019uncertainty}.
\item Communication as a process: model communication as a process. For example, the Lasswell model\cite{qiming2021strategy}, Shannon \& Weaver model~\cite{zhou2007ontology}, Westley \& MacLean's model~\cite{vitkova2021approach}, and Speech Act Theory~\cite{9199096}.
\end{itemize}

In the analyzed literature, non-systemic approaches predominate, focusing on a particular stage of a piece of digital false information's life (413 articles on detection in diffusion or consumption). The results reveal 5 stages of communication: Creation~\cite{oehmichen2019not}, Diffusion~\cite{hui2018hoaxy}, Consumption~\cite{starbird2019disinformation}, Decay~\cite{hui2018hoaxy}, and Transformations~\cite{starbird2019disinformation, shin2018diffusion, webb2016digital}. The research characterizes each stage considering aspects such as the motivations of the stakeholders involved in the creation of digital false information \cite{jang2018computational}, the activities involved in creating digital false information (preparation~\cite{oehmichen2019not, hussain2018analyzing}, elaboration~\cite{mccarthy2018let}, dissemination of false content~\citep{boshmaf2013design, traylor2017psyop, raynal2010small}), and the ethical intentions of users in sharing digital false information~\cite{webb2016digital}.

Table~\ref{tab:carac_estag} presents characteristics of each stage.

\begin{table}[H]
\centering
\caption{Characteristics of stages of the digital false information phenomenon.}
\label{tab:carac_estag}
\footnotesize
\renewcommand{\tabularxcolumn}[1]{m{#1}}
\begin{tabularx}{\textwidth}{>{\hsize=0.25\hsize\linewidth=\hsize\centering\arraybackslash}X>{\hsize=1.75\hsize\linewidth=\hsize\raggedright\arraybackslash}X}
Stage & Characteristics \\ \hline
Creation &
\begin{itemize}[noitemsep, label={}]
\item Can be a collaborative or solitary activity~\cite{hussain2018analyzing};
\item Collaborative activities require planning and coordination~\cite{hussain2018analyzing};
\item Sub-activities: preparation~\cite{oehmichen2019not}, elaboration~\citep{mccarthy2018let}, and dissemination~\cite{traylor2017psyop, boshmaf2013design};
\item Events such as crises, natural disasters, and terrorist attacks are linked to the emergence and virality of digital false information~\cite{oh2010exploration};
\end{itemize}
\\\hline
Diffusion &
\begin{itemize}[noitemsep, label={}]
\item The communication format affects audience interaction~\cite{khodabakhsh2018taxonomy};
\item Communication can be synchronous or asynchronous~\cite{ zhou2007ontology}, bidirectional or unidirectional~\cite{neumann1996disinformation};
\item Asynchronous limit the communication control of the audience~\cite{Kumar2016psychometric};
\end{itemize}
\\ \hline
Consumption &
\begin{itemize}[noitemsep, label={}]
\item Aspects of the message affect the audience's perception of risk, such as message length~\cite{kumar2016disinformation, zannettou2019web}, links~\cite{aigner2017manipulating}, allowing images~\cite{taylor2018democratic}, layout~\cite{leite2019waste, aigner2017manipulating}, and colors~\cite{taylor2018democratic};
\item The layout of the message directs the reader's attention~\cite{shen2019media};
\item Social aspects such as relationships (authorities, affective proximity, etc.) and beliefs (political, religion, etc.) influence consumption~\cite{rath2019evaluating, berghel2017lies};
\item Communities may have unhealthy information consumption habits, such as consuming and sharing without verifying\cite{kovic2016consuming};
\item Human factors are susceptible to manipulation, such as memory~\cite{larson2019applying}; can be exploited by logical fallacies and cognitive biases~\cite{li2012argument};
\end{itemize}
\\ \hline
Reiteration &
\begin{itemize}[noitemsep, label={}]
\item After the initial dissemination, digital false information can migrate between different digital communication platforms~\citep{janicka2019cross, del2017news}, or between different groups on the same platform~\cite{melo2019whatsapp};
\item Organic content changes, with the audience appropriating the message, or artificial, with \textit{bots} recreating or redirecting content~\cite{starbird2019disinformation, yu2018silent};
\item Stakeholders may make adjustments to avoid detection~\cite{bandeli2018analyzing}, or promote corrected messages in a new format~\cite{finn2015spread, tang2017echo}.
\end{itemize}
\\ \hline
\end{tabularx}

\end{table}

\subsubsection{Discussions}

Research tends to approach the phenomenon of digital false information in a fragmented manner, with a limited scope focused on some stage of a piece of digital false information's lifecycle. For example, studies on digital false information detection often concentrate on diffusion and consumption, considering little or no aspect of the creation of digital false information. In this case, the creation of digital false information targets a specific audience (who communicates something, or communicates to someone, especially when deliberately disinforming). As a result, there is a limited understanding of the phenomenon that restricts the identification of interconnected aspects in the stages of a piece of digital false information's life.

Although the approaches contribute to understanding the phenomenon, they present limitations in terms of completeness, comprehensiveness, and interconnections of the sociotechnical aspects involved in a case of digital false information. The models are not comprehensive enough to support the study of deliberate and accidental digital false information~\footnote{A challenging characteristic of the phenomenon is the analysis of intentionality and deliberation. Users often make mistakes and create and spread digital false information accidentally~\cite{watney2018legal}, which poses a challenge for classifiers and even more so for responsible digital false information mitigation.}, as they are based on assumptions related to intentionality. The life cycle of deliberate digital false information focuses on the perspective of the creator or the consumer separately. Models that address diffusion tend to sideline stakeholders~\cite{fernandez2018online}. Furthermore, models that consider communication as a process, although they recognize intentionality as a crucial element in the constitution of a message~\cite{fiske2010introduction} (such as the Speech Act Theory~\cite{9199096}), are limited by adopting the perspective of the interlocutor as having the greatest impact in the context of message interpretation, assuming that the consumer's perspective is the same as the speaker's~\cite{sbisa1980models}. This limits the model's ability to represent reality, especially from the perspective of the audience that consumes the digital false information, and it disregards stakeholders who control the communication channel. These limitations represent barriers to understanding interconnected aspects in the stages of a piece of digital false information's life and to the "generalization"\footnote{With this term, the author refers to the transferability of knowledge generated by \citep{oulasvirta2016hci}: "The number of users, tasks, and contexts to which the solution can be applied; qualitatively new contexts to which the solution can be applied." .} of the models.

The way these models approach the phenomenon limits the ability to analyze a case of digital false information comprehensively (considering sociotechnical aspects and the relationship between the stages of a digital false information's lifecycle) and flexibly (allowing exploration of different sociotechnical aspects at different levels of abstraction). There is a tendency to address digital false information in a fragmented manner, focusing on the analysis of quantifiable attributes of a specific stage of digital false information's life or in a structured manner through models that generalize assumptions about qualitative aspects of digital false information. Therefore, it is necessary to comprehensively understand digital false information as a sociotechnical phenomenon, where social and technical aspects are interconnected in the stages of a piece of digital false information's life.

There is room for investigating sociotechnical and systemic approaches to assist in the analysis of digital false information, especially considering stakeholders as a central element of the phenomenon. Due to the complexity of the phenomenon, it is necessary to support intervention designers in understanding the sociotechnical nature of digital false information. A framework of epistemic tools is needed to help comprehensively identify aspects of stakeholder involvement in digital false information communication.

\vspace{-0.5em}
\section{Threats to Mapping Validity}

The following types of validity are considered based on~\cite{petersen2015guidelines}: descriptive validity, interpretative validity, and theoretical validity. Given the general nature of this work, threats to generalization~\cite{petersen2015guidelines} were considered to have a minimal impact on the study. To reduce threats to descriptive validity, which concerns how we conduct documentation, data collection forms were used to support the data selection and extraction process, based on~\cite{petersen2015guidelines, wieringa2006requirements}. They were designed to maintain rigor in coding and obtain objective records. In the case of threats to theoretical validity, determined by confounding factors such as bias in study selection~\cite{petersen2015guidelines}, we adopted a protocol and review strategies after each filter. Studies not accessible through Capes were not included; hence, relevant research may have been overlooked. Regarding interpretative validity---the credibility of interpretations~\cite{petersen2015guidelines}---only the author discusses the conclusions, which may have a relevant impact. However, I made an effort to record and clarify how I derived the conclusions to enhance both my perception and research capability.

\section{Final Remarks}
This article summarizes the findings of a systematic literature review in the field of Computer Science that explores digital false information. The review reveals a tendency to analyze the phenomenon in a segmented manner, focusing on specific stages of the false information lifecycle that are often unrelated. Some studies adopt restrictive communication models, with a generic focus on diffusion as a process, addressing the perspective of specific stakeholders, usually the consumer or the digital false information creator. Furthermore, addressing the phenomenon as sociotechnical proves to be a complex task, with a range of relevant elements at different levels of abstraction requiring methodologies that integrate various artifacts, methods, and theories, demanding knowledge and experience in quantitative and qualitative analysis techniques.

Moreover, there are limited considerations of interrelated sociotechnical aspects, with predetermined groups of aspects used in other studies. There are difficulties in relating aspects at different levels of abstraction, which minimally encourages the identification and study of new types of aspects and their interrelations. Additionally, few studies propose methods to analyze a digital false information case in a situated manner, seeking sociotechnical aspects linked to possible impacts or influencing factors for a well-defined context.

There is a need for research into sociotechnical and systemic approaches that aid in digital false information analysis, especially considering stakeholders as a central part of the phenomenon. Due to the complexity of the phenomenon, it is necessary to support stakeholders in investigating and mitigating it, understanding the sociotechnical nature of digital false information, assisting in identifying sociotechnical aspects at different levels of abstraction, and identifying their interrelations. Therefore, a sociotechnical and systemic approach is required for digital false information analysis, which would equip stakeholders with the tools to analyze digital false information cases comprehensively. 

This research makes several contributions including: a classification of mitigation strategies that are suitable for building a catalog of intervention types, a diverse set of relevant socio-technical aspects and challenges that are organized into six semiotic dimensions (which are drawn from Organizational Semiotic), a discussion on the interplay between sociotechnical aspects to deepen the understanding of complex interrelationships among technical and social aspects of the phenomenon, and a classification of digital false information types based on a sociotechnical perspective.




\bibliographystyle{plainnat}

\bibliography{sbc-template}

\end{document}